\documentclass[aps,prb,twocolumn,letterpaper,showpacs,10pt]{revtex4-1}
\usepackage{graphicx,amsmath,mathrsfs}
\usepackage{amssymb}
\usepackage{amsthm}
\usepackage{bm}
\usepackage{float}
\usepackage{epsfig}

\usepackage{color}

\def\bc{\begin{center}}
\def\ec{\end{center}}

\newcommand{\bs}[1]{\boldsymbol{#1}}

\newcommand{\ket}[1]{\left|#1\right\rangle}
\newcommand{\bra}[1]{\left\langle#1\right|}
\newcommand{\braket}[2]{\bigl\langle#1\bigl|\bigr.#2\bigr\rangle}

\renewcommand {\ec}{\eta_{\gamma}}

\def\bra#1{\left\langle#1\right|}
\def\ket#1{\left|#1\right\rangle}
\def\braket#1#2{\left\langle #1\right|\left.#2\right\rangle}
\def\vk{{\bf k}}
\def\vr{{\bf r}}

\begin{document}
\title{Pseudopotential Formalism for Fractional Chern Insulators}
\author{Ching Hua Lee${}^1$}
\author{Ronny Thomale${}^{2,3}$}
\author{Xiao-Liang Qi${}^1$}
\affiliation{${}^1$Department of Physics, Stanford University, Stanford, CA 94305, USA}
\affiliation{${}^2$ Institut de th\'eorie des ph\'enom\`enes physiques, \'Ecole Polytechnique F\'ed\'erale de Lausanne (EPFL), CH-1015 Lausanne}
\affiliation{${}^3$Institute for Theoretical Physics and Astrophysics, University of W\"urzburg, D 97074 W\"urzburg}

\date{\today}
\begin{abstract}

  Recently, generalizations of fractional quantum Hall (FQH) states known as fractional quantum
  anomalous Hall or, equivalently, fractional Chern insulators states have been
  realized in lattice models. Ideal wavefunctions such as the Laughlin wavefunction, as well as their corresponding
  trial Hamiltonians, have
  been vital to characterizing FQH phases. The Wannier function representation of fractional Chern
  insulators proposed in [X.-L. Qi, Phys. Rev. Lett. {\bf 107}, 126803]
  defines an approach to generalize these concepts to fractional Chern insulators. In this paper, we apply
  the Wannier function representation to develop a systematic
  pseudopotential formalism for fractional Chern insulators. The family
  of pseudopotential Hamiltonians is defined as the set of projectors onto
  asymptotic relative angular momentum components which forms an
  orthogonal basis of two-body Hamiltonians with magnetic translation
  symmetry. This approach serves both as an expansion tool for
  interactions and as a definition of positive semidefinite
  Hamiltonians for which the ideal fractional Chern insulator
  wavefunctions are exact nullspace modes. We compare the short-range
  two-body pseudopotential expansion of various fractional Chern
  insulator models at filling $\mu = 1/3$ in phase regimes where a
  Laughlin-type ground state is expected to be realized. We also
  discuss the effect of inhomogeneous Berry curvature which leads to
  components of the Hamiltonian that cannot be expanded into
  pseudopotentials, and elaborate on their role in determining low
  energy theories for fractional Chern insulators. Finally, we generalize our Chern
  pseudopotential approach to interactions involving more than two bodies with the goal
  of facilitating the identification of non-Abelian fractional Chern
  insulators.

\end{abstract}
\maketitle

\section{Introduction}
\label{sec:intro}

The field of topological insulators (TIs) is currently witnessing enormous interest
in condensed matter~\cite{HasanKane,xlreview,moore2010}. The predecessor of TIs is the
integer quantum Hall effect (IQHE)~\cite{klitzing-80prl494} realized
in a two-dimensional electron gas with a strong perpendicular magnetic
field. Similar to other TIs discovered more recently, the IQHE is a
gapped state of matter characterized by topologically robust edge
states and a bulk topological invariant known as the Chern number or
the Thouless-Kohmoto-Nightingale-den Nijs (TKNN)
number\cite{thouless82-prl405}. The IQH state has been generalized to
lattice models without orbital magnetic field\cite{haldane88prl2015},
which are named as quantum anomalous Hall (QAH) states or Chern
insulators (CI). QAH states have been proposed in realistic
materials\cite{qi2006b,liu2008B,yu2010}.
 In 1983, the fractional quantum Hall effect (FQHE) was discovered in
 systems with a fractionally filled Landau
 level\cite{tsui-82prl1559,laughlin83prl1395}. Since then, the FQHE has become a paradigmatic example
of a topologically ordered phase~\cite{wen-89prb7387} where interactions exhibit non-perturbative roles.
Numerous fundamental developments evolved out of this direction, such as the
concept of non-Abelian statistics which forms the foundation of topological
quantum computation~\cite{Moore-91npb362,nayak-08rmp1083}. One fundamental difference from IQHE is that the flatness of the
Landau level and its associated freezing of kinetic energy appear to be necessary conditions for the FQHE state to be energetically preferable.

This immediately provokes the question of whether a fractional Chern
insulator (FCI) can be realized, i.e. a lattice version of the FQHE
without an external magnetic field. Taking a CI such as Haldane's
honeycomb model~\cite{haldane88prl2015} as a natural starting point, the task then is to drive
the system into the flat band limit where the chemical potential lies
within this band, e.g. at fractional one third filling, which is well
separated from the other bands and hence accomplishes a FQHE-type
lattice scenario. (Unlike the quantum Hall case, the FCI filling is
not given by the ratio of electrons over magnetic flux quanta, but the
chemical potential of the lattice model.) Different groups have
recently independently pursued this direction, proposing FCI models on
the honeycomb, kagome, square, and checkerboard
lattice~\cite{tang-11prl236802,neupert-11prl236804,sun-11prl236803}.
 In different ways, the flattening of the Chern band can be
accomplished through geometric frustration (e.g. long-range
hopping)~\cite{tang-11prl236802,neupert-11prl236804}, multi-band
effects~\cite{sun-11prl236803}, and multi-orbital
character~\cite{verderbos-11prl116401}. While the $s$ and
$p$-type orbitals in previous candidates materials for topological insulators would assume only moderate interactions from small
hybridizations, $d$-orbital-type systems provide an arena for both strong correlations and topological band
structures~\cite{xiao-11ncomm596}.
First numerical investigations of
the FCI phases on a torus at one third band filling found indications
of a three-fold topologically degenerate ground state separated from the other energy levels by a
gap, where the flux insertion showed
level crossings with no level repulsion between them, and the Chern
numbers of these many-body ground states found to be $1/3$
each~\cite{sheng-11ncomm389,neupert-11prl236804}.
While this
already gives a strong hint that a Laughlin-type fractional Chern
phase might be realized, this does not yet completely rule out a
competing charge density wave (CDW) state at this filling, which can
show similar fractional Chern numbers in the ground states, level
degeneracy, and a gap. Further evidence against a CDW
state, however, has been found by finite size scaling, entanglement measures, and the distribution of ground state momenta as a
function of cluster size\cite{regnault-11prx021014}. Compared to
the FQHE for which the joint perspective of energy and entanglement measures
generally gives a consistent and complementary picture, the current
stage of FCI models particularly calls for further investigation.

In this paper, we focus on further developing the
understanding of FCI phases from the perspective of energetics and
interactions. In general, FCIs involve different scales such as the
kinetic bandwidth of the fractionally filled Chern band, the gap
separation from other bands, as well as the magnitude and range of
interactions. Even if we assume a conventional FQHE-type parameter
window where inter-band scattering is neglected and the bandwidth of
the fractional Chern band is assumed small versus the interaction
strength, a crucial complication of FCI models is the inhomogeneous
Berry curvature which has no quantum Hall analogue. As a signature of this difference, the
Platzman-Girvin-MacDonald algebra~\cite{girvin-86prb2481} of the
lowest Landau level (LLL) can only be mapped to the lattice Chern density
operators in the continuum limit as well as for homogeneous Berry
curvature~\cite{parameswaran-cm1106}, from where Hamiltonian theories
can be constructed~\cite{ramamurthy}. However, an exact one-to-one mapping between FQH and FCI states with Chern number $C=1$ has been established by the Wannier state representation of Chern insulators\cite{qi11prl126803} despite of the inhomogeneous Berry curvature. (One-dimensional) Wannier states are single particle states which are localized in real space in one spatial direction (such as $x$), but are momentum eigenstates in the orthogonal direction $y$. Different Wannier states are related by translation in $x$ direction, and all Wannier states form a complete basis of the single particle Hilbert space defined by a non-degenerate energy band. The exact mapping between FQH and FCI is obtained by mapping Landau level wavefunctions in the Landau gauge to the Wannier states in FCI. More details of this mapping will be reviewed in Sec. \ref{sec:wannier}. From the Wannier state representation, we learn that the effect of inhomogeneous Berry curvature is absent if we consider a special Hamiltonian obtained by mapping a FQH Hamiltonian to the FCI system. In other words, the effect of inhomogeneous Berry curvature in an FCI strongly depends on the interaction Hamiltonian. In two recent works\cite{wu2012,gunnar}, the Wannier state representation and its further improvement has been investigated in both fermionic and bosonic FCI systems. In the bosonic $\nu=1/2$ FCI, the wavefunction proposed by the Wannier state representation has a high overlap with the exact ground state wavefunction obtained by exact diagonalization\cite{gunnar}. In the fermionic $\nu=1/3$ case, such a high overlap is also achieved given that the Wannier states are  modified\cite{wu2012}. Therefore the validity of the Wannier state representation has been demonstrated at least in those simplest FCI states.

An important part of this paper will concern the development of a pseudopotential (PP) formalism for fractional Chern
insulators. Previously, PPs have been established in the context of
FQHE~\cite{haldane83prl605}. PPs are partial wave
expansions of the Coulomb interactions. The resulting expansion
quantum number is the relative angular momentum $m$ of two particles
(with $m$ even for bosons and odd for
fermions) where the expansion coefficients $V_m$ denote the energy
penalty of two particles having a relative angular momentum of $m$. In
the same way the Landau level wave functions were used as a basis
for defining such an expansion in FQHE, we now employ a similar
construction for the FCI Wannier functions~\cite{qi11prl126803}.  PPs
proved extremely useful in FQHE not only to give a universal
classification of different interaction profiles, but also to obtain
an adequate description of general FQHE phase diagrams. Furthermore,
many paradigmatic FQH wave functions are exact ground states of
certain PP Hamiltonians for which representative finite
size studies would be more accurate to resolve their universal
properties than for a generic interaction
scenario~\cite{haldane-85prb2529,simon-07prb075318,mcs}. In adapting
this concept to FCI models, we can hope for a similarly promising
route to a deeper and more universal understanding of FCI
Hamiltonians.

The paper is organized as follows. In Section~\ref{sec:wannier}, we
review and expand the description of the Wannier state representation
of fractional Chern bands. This lays the foundation for the definition
of FCI PPs, for which we first review the PP formalism
for FQHE and subsequently develop the FCI formulation of PPs in
Section~\ref{sec:pp}.  In Section~\ref{sec:model}, we apply our PP
formalism to different FCI models and their interaction profiles. We
find that, as a direct consequence of inhomogeneous Berry curvature,
there is a portion of the interactions which cannot be expanded in
PPs. These observations will be analysed in detail in Section~\ref{sec:CMMT}.
There, we classify what type of center-of-mass (CM) breaking
and magnetic-translation-group (MT) breaking scattering elements appear in
FCI interactions, and how these symmetries are potentially reemergent
in an effective low energy theory of the
problem~\cite{haldane85prl2095,bernevig-12prb075128}. In
Section~\ref{sec:mbpp}, we generalize
the FCI PP principle to many-body interactions, which enables us to define exact Hamiltonians
for non-Abelian FCI phases. In Section~\ref{sec:concl}, we conclude
that the pseudopotential formalism establishes a suitable platform
to further investigate and analyze new states of matter in FCIs.

\section{Wannier state representation of fractional Chern bands}
\label{sec:wannier}

In this section, we review the Wannier state representation of FCIs
proposed in Ref. \onlinecite{qi11prl126803}. The idea of this approach is to
find a suitable single-particle basis, the one-dimensional (1D)
Wannier state basis, and to use this basis to establish an exact mapping
between FCI and FQH states. While a cylindrical geometry was employed in Ref. \onlinecite{qi11prl126803}, the discussion can also be formulated on the torus geometry\cite{yu2011} which we use in the following. (The torus formulation of the Wannier state representation has also been investigated independently in Ref. \cite{wu2012,gunnar}.)

Consider a band insulator with the Hamiltonian
\begin{eqnarray}
H=\sum_{i,j,\alpha,\beta}c_{i\alpha}^\dagger h_{ij}^{\alpha\beta}c_{j\beta},
\end{eqnarray}
with $i,j$ being the site indices of a two-dimensional lattice with periodic boundary conditions and $\alpha,\beta=1,2,..,N$ labeling internal states in each unit cell such as orbital and spin states. With translation symmetry $h_{ij}^{\alpha\beta}=h_{{\bf r}_j-{\bf r}_i}^{\alpha\beta}$, the Hamiltonian can be written in momentum space as
\begin{equation}
H=\sum_\vk c_{\vk \alpha}^\dagger h_\vk^{\alpha\beta}c_{\vk
  \beta},\label{Hk}
\end{equation}
with
\begin{eqnarray}
c_{\vk\alpha}&=&\frac1{\sqrt{L_xL_y}}\sum_{i}c_{i\alpha}e^{-i\vk\cdot{\bf r}_i},\nonumber\\
h_{{\bf r}_j-{\bf r}_i}^{\alpha\beta}&=&\frac1{L_xL_y}\sum_\vk h_{\vk}^{\alpha\beta}e^{-i{\vk\cdot(\vr_j-\vr_i)}}.\nonumber
\end{eqnarray}
We use $L_x,L_y$ to denote the number of lattice sites in $x$ and $y$ direction, respectively. The momentum $\vk$ takes values of $\left(\frac{2\pi n_x}{L_x},\frac{2\pi n_y}{L_y}\right)$, with $n_x=1,2,...,L_x,~n_y=1,2,...,L_y$ integers. The Hamiltonian matrix $h_\vk$ can be diagonalized to obtain the eigenstates
\begin{eqnarray}
h_\vk\ket{n,\vk}=E_{n\vk}\ket{n,\vk},~n=1,2,..,N.
\end{eqnarray}
We are interested in the system with a lowest energy band $E_{1\vk}$ occupied, and a gap separating this band from all other bands. Since only the lowest band will be involved, we will denote $\ket{1,\vk}$ by $\ket{\vk}$ for simplicity.

In the thermodynamic limit $L_x,L_y\rightarrow \infty$, ${\bf k}$ is a
good quantum number and the Berry's phase gauge field ${\bf
  a}=-i\bra{\vk}\nabla_\vk\ket{\vk}$ can be defined, which determines
the first Chern number as the flux of the gauge field in the Brillouin
zone: $C_1=\frac1{2\pi}\int_{BZ} d^2\vk \nabla\times{\bf a}$. For the
realization of FCIs, we are interested in a band with $C_1\neq 0$. More
specifically, in this paper we will focus on $C_1=1$
systems. Moreover, for finite $L_x,L_y$, it is necessary to generalize
the definition of a Berry's phase gauge field and Chern number to
the case of $\ket{\vk}$ with a discrete $\vk$ variable.

We start from the definition of 1D Wannier states
\begin{eqnarray}
\ket{W_{nk_y}}=\frac1{\sqrt{L_x}}\sum_{k_x}e^{-ik_xn}e^{i\varphi_\vk}\ket{\bf k},\label{eq:Wannier}
\end{eqnarray}
which is a Fourier transform of $\ket{\vk}$ in the $x$-direction, but which remains an eigenstate of $k_y$. Since the state $\ket{\vk}$ is only determined by the Hamiltonian up to a phase, the phase factor $e^{i\varphi_\vk}$ is not pre-determined. As was discussed in Refs. \onlinecite{kivelson1982,qi11prl126803}, the phase ambiguity can be fixed by defining the projected position operator $\hat{x}=PxP$ with
$P=\sum_\vk\ket{\vk}\bra{\vk}$, the projection operator to the occupied
band, and $x=\sum_{i,\alpha}x_i\ket{i,\alpha}\bra{i,\alpha}$ the
position operator. However, in a system with periodic boundary
conditions in $x$-direction, it will be slightly more problematic to apply
this definition of $x$ due to the dependence on the choice of the boundary site. As pointed out in Ref.~\onlinecite{yu2011}, this problem can be resolved by defining a unitary operator
\begin{eqnarray}
X=\exp\left[ix\frac{2\pi}{L_x}\right]=\exp\left[i\sum_{i}\frac{2\pi x_i}{L_x}\sum_{\alpha}\ket{i\alpha}\bra{i\alpha}\right].
\end{eqnarray}
This definition preserves the periodicity $x\rightarrow x+L_x$. The eigenstates of this operator are the states localized on a given site $n$ in $x$-direction. We define the projected operator
\begin{eqnarray}
\hat{X}=PXP.
\end{eqnarray}
In momentum space, $\bra{\vk}\hat{X}\ket{\vk'}=\bra{\vk}X\ket{\vk'}$. It is easy to see that $X$ shifts the momentum $k_x$ by $2\pi/L_x$, since
\begin{eqnarray}
\ket{\vk}&=&\frac1{\sqrt{L_xL_y}}\sum_{i,\alpha}u_{\vk \alpha}e^{ix_ik_i}\ket{i,\alpha}\nonumber,\\
X\ket{\vk}&=&\sum_{i,\alpha}u_{\vk \alpha}e^{ix_i\left(k_x+\frac{2\pi}{L_x}\right)}\ket{i,\alpha},
\end{eqnarray}
with $u_{\vk \alpha}=\sqrt{L_xL_y}\braket{0,\alpha}{\vk}$ the periodic part of the Bloch wave function.
Therefore, the only nonzero matrix element of $\bra{\vk}X\ket{\vk'}$ is
\begin{eqnarray}
F_{k_xk_y}&\equiv&\bra{k_x+2\pi/L_x,k_y}X\ket{k_x,k_y}
\nonumber\\
&=&\sum_\alpha u_{k_x+\frac{2\pi}{L_x},k_y,\alpha}^*u_{k_x,k_y,\alpha}.
\end{eqnarray}
In the subspace of states with given $k_y$, the matrix of $\hat{X}$ in momentum representation is
\begin{eqnarray}
\hat{X}=\left(\begin{array}{ccccc}0&&..&&F_{2\pi}\\F_{2\pi/L_x}&0&&&\\&F_{4\pi/L_x}&&&..\\
..&&..&&\\&&&F_{(L_x-1)\pi/L_x}&0\end{array}\right),
\end{eqnarray}
where the omitted index $k_y$ is the same for all states.

\begin{figure}
\begin{minipage}{0.7\linewidth}
\includegraphics[width=\linewidth]{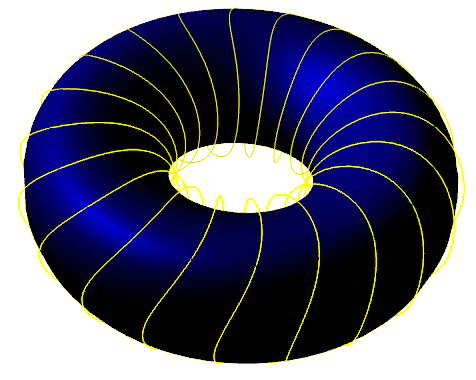}
\end{minipage}
\caption{(Color online) Shift of the Wannier polarization as $k_y$
  varies. The azimuthal direction represents the real space $x$ while
  the poloidal direction represents the periodic domain of $k_y$. For
  any full rotation along $y$, i.e. $k_y\rightarrow k_y + 2\pi$, the
  Wannier state shifts by $x\rightarrow x+1$.}
\label{fig:wann_shift}
\end{figure}

In the thermodynamic limit $L_x\rightarrow \infty$, $F_{k_xk_y}\simeq 1-i\frac{2\pi}{L_x}a_{x}(\vk)$, with $a_x$ the $x$ component of the Berry's phase gauge field. For finite $L_x$, $\left|F_{k_xk_y}\right|$ should be close to but not exactly equal to $1$. Therefore, $\hat{X}$ is an approximately unitary matrix. To define the maximally localized Wannier states, we deform the $\hat{X}$ operator to a unitary operator by defining
\begin{eqnarray}
F_{k_xk_y}&=&\left|F_{k_xk_y}\right|e^{-iA_{k_xk_y}}\nonumber\\
\bar{X}&=&\left(\begin{array}{ccccc}0&&..&&e^{-iA_{2\pi}}\\e^{-iA_{2\pi/L_x}}&0&&&\\&e^{-iA_{4\pi/L_x}}&&&..\\
..&&..&&\\&&&e^{-iA_{(L_x-1)\pi/L_x}}&0\end{array}\right).\nonumber\\
\end{eqnarray}
Here, the index of the rows and columns are
$k_x=0,\frac{2\pi}{L_x},...,2\pi-\frac{2\pi}{L_x}$. The eigenstates of
the $\bar{X}$ operator form an orthogonal complete basis. Due to the simple form of $\bar{X}$ in momentum space, one can prove that the eigenstates of $\bar{X}$ are Wannier states defined in Eq. (\ref{eq:Wannier}), with the phase $\varphi_\vk$ defined by
\begin{eqnarray}
\ket{W_{nk_y}}&=&\frac1{\sqrt{L_x}}\sum_{k_x}e^{-ik_xn}e^{i\varphi_\vk}\ket{\bf k},\nonumber\\
\varphi_\vk&=&-\sum_{0\leq p_x< k_x}A_{p_xk_y}-k_xP_x(k_y),\label{eq:MLWF}
\end{eqnarray}
with
\begin{equation}
P_x(k_y)=-\frac1{2\pi}\sum_{0\leq p_x<2\pi}A_{p_xk_y}.\nonumber
\end{equation}
This definition is periodic in $k_x\rightarrow k_x+2\pi$. The corresponding eigenvalues are
\begin{eqnarray}
\bar{X}\ket{W_{nk_y}}=e^{i\frac{2\pi}{L_x}\left(n+P_x\right)}\ket{W_{nk_y}}.
\end{eqnarray}
Therefore, we see that the center-of-mass (CM) position of the state
$\ket{W_{nk_y}}$ is shifted by $P_x$ away from the lattice site
position $n$. This fact indicates that $P_x(k_y)$ has the physical
meaning of charge polarization\cite{kingsmith1993}. In the large $L_x$ limit, $A_{\vk}\rightarrow \frac{2\pi}{L_x}a_x$ and $P_x(k_y)=-\frac1{2\pi}\int_0^{2\pi}a_xdk_x$. Since $P_x(k_y)$ is a $U(1)$ phase for each $k_y$, one can define its winding number when $k_y$ goes from $0$ to $2\pi$:
\begin{equation}
C_1=\sum_{n=1}^{L_y}\frac{i}{2\pi}\log\left(e^{i2\pi\left[P_x(2\pi n/L_y)-P_x(2\pi(n-1)/L_y)\right]}\right).
\end{equation}
The $\log$ function is defined to take the value of the phase
difference $2\pi\left[P_x(2\pi n/L_y)-P_x(2\pi(n-1)/L_y)\right]$ in
the region of $[-\pi,\pi)$. As long as $L_y$ is not too small so that
$P_x(k_y)$ would not jump by an integer between two neighboring $k_y$ values, the $C_1$ obtained agrees with the Chern number in the large $L_y$ limit.

\begin{figure}
\begin{minipage}{0.99\linewidth}
\includegraphics[width=\linewidth]{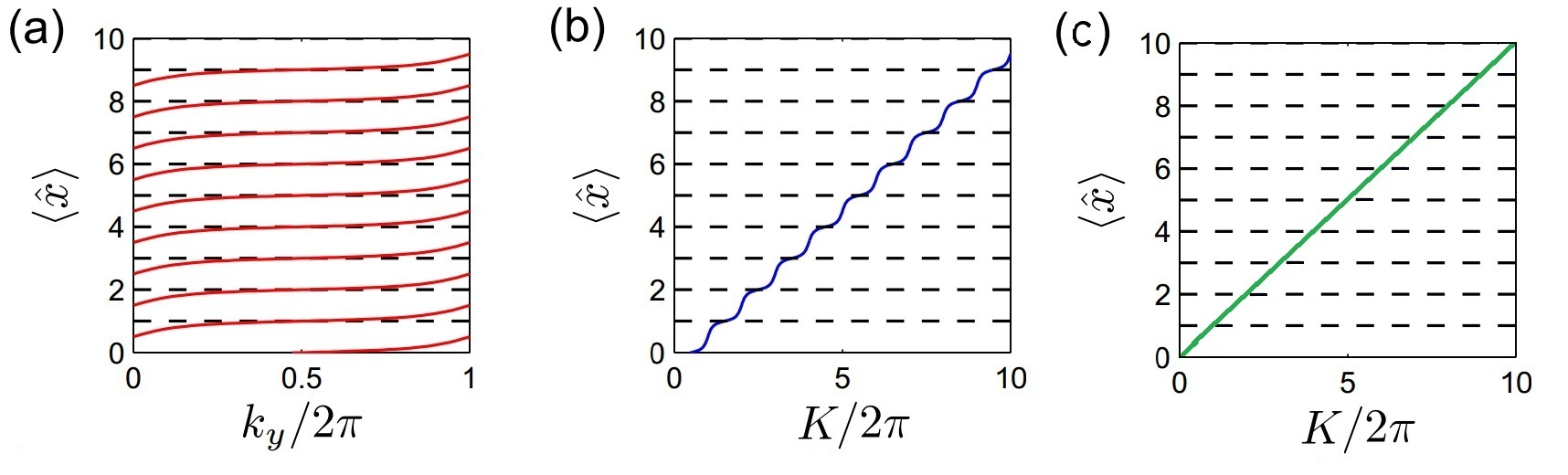}
\end{minipage}
\caption{(Color online) Wannier center evolution of the Dirac model used in Ref. \onlinecite{qi11prl126803} from $x=0$ to
  $x=10$ as a function of (a) $k_y/2\pi$ and (b) of the extended wave vector $K/2\pi$. From (a) to (b), it becomes visible how the evolution
of $x$ changes
as a function of $K$, suggesting its similarity to Landau level wave
functions (c).}
\label{fig:wann}
\end{figure}

For $C_1\neq 0$, the Wannier states $\ket{W_n(k_y)}$ have a ``twisted
boundary condition" in $k_y$, since $P_x(k_y+2\pi)=P_x(k_y)+C_1$, such
that $\ket{W_n(k_y+2\pi)}=\ket{W_{n+C_1}(k_y)}$. As is illustrated in
Fig.~\ref{fig:wann_shift}, for $C_1=1$ the Wannier state CM
position $x_n(k_y)=n+P_x(k_y)$ forms a helical curve on the parameter
space torus of  $x,k_y$. The key observation which enables the Wannier state representation of FCI is the fact that the twisted boundary condition allows to label all Wannier states in such a 2D system by a 1D parameter. If we define
\begin{eqnarray}
\ket{W_{k_y+2\pi n}}=\ket{W_{n,k_y}},~\text{for~}k_y\in[0,2\pi),
\end{eqnarray}
it yields $\ket{W_K}$, with $K=k_y+2\pi n$ a continuous function of $K$ (in the large $L_y$ limit). The CM position $x_K=\bra{W_K}\bar{X}\ket{W_K}=e^{i\frac{2\pi}{L_x}
\left(n+P_x(k_y)\right)}$ for $K\in[2\pi n,2\pi(n+1))$ is continuous
in $K$ and satisfies $x_{K+2\pi}=x_{K}+1$. In this sense, $x_K$ increases
linearly with $K$ (Fig.~\ref{fig:wann}).

Due to this behavior of $\ket{W_K}$, an exact mapping can be defined
between the Wannier states in the $C_1=1$ FCI and the LLL states in a
FQH problem. Consider a spinless fermion with the Hamiltonian
$H=\frac1{2m}({\bf p}-{\bf A})^2$ on a torus of the size $L_xl_B\times
L_yl_B$ with the uniform perpendicular magnetic field
$\nabla\times{\bf A}=B=2\pi/l_B^2$. The total number of flux is
$N_\Phi=L_xL_y$, so that the LLL contains the same dimension of
Hilbert space as the lattice model discussed above on a lattice of the
size $L_x\times L_y$. In a gauge choice $A_x=0,~A_y=Bx$, the LLL  wave functions have the form
\begin{eqnarray}
\psi_K(x,y)&=&\frac1{\sqrt{2\pi l_B^2}}\sum_{n\in {\rm Z}}e^{iKy/l_B-\pi\left(\frac{x}{l_B}-\frac{K}{2\pi}-nL_x\right)^2}\nonumber\\
&\equiv &\frac1{\sqrt{2\pi l_B^2}}e^{iKy/l_B-\pi\left(\frac{x}{l_B}-\frac{K}{2\pi}\right)^2}\nonumber\\
& &\cdot\vartheta\left(-i{L_x}\left(\frac{x}{l_B}-\frac K{2\pi}\right)
  \vert {iL_x^2}\right)\label{torusLLWF}
\end{eqnarray}
with $\vartheta(z \vert \tau)$ the Jacobi theta function~\cite{Abra} which are
appropriate superpositions of the cylindric wave functions.

Notice that we have defined the momentum $K$ slightly differently from
 the usual definition used in Ref.~\onlinecite{lee-04prl096401,qi11prl126803} and in the subsequent sections, so that here $K$ is dimensionless and given by $K=\frac{2\pi}{L_y}n,~n\in{\mathbb{Z}}$ on the $L_xl_B\times L_yl_B$ torus. This definition leads to identical results as the usual definition used later if we replace the $l_B$ here by $\sqrt{2\pi}l_B$.

For $L_x\gg 1$, this wave function is a Gaussian function around the CM position $x_K=K\frac{l_B}{2\pi}$. Denoting $\ket{\psi_K}$ as the state corresponding to wave function $\psi_K(x,y)$, one can define a unitary mapping between the Hilbert spaces of the Landau level and the lattice $C_1$ Chern insulator:
\begin{eqnarray}
f:~{\rm H}_{\text{CI}}&\longrightarrow& {\rm H}_{\text{LLL}},\nonumber\\
f\left(\ket{W_K}\right)&=&\ket{\psi_K},\label{mapFQHFCI}
\end{eqnarray}
with ${\rm H}_{\text{CI}}$ and ${\rm H}_{\text{LLL}}$ denoting the Hilbert spaces of the CI and LLL, respectively.
Such a mapping preserves the continuity in $K$ and also the
topological properties of $\ket{W_K}$ and $\ket{\psi_K}$, i.e., their
winding while momentum $K$ is increased. Using the reverse map
$f^{-1}$, the many-body states of the LLL, such as
Laughlin states and other FQH states defined in the LLL, can all be mapped to corresponding states in the
FCI. Similarly, a Hamiltonian $H$ of a FQH system can also be mapped
to a corresponding Hamiltonian $H_{\text{FCI}}=f^{-1}Hf$. The main
purpose of the current paper is to study the Hamiltonians
$H_{\text{FCI}}$ which are mapped from the PP Hamiltonians in the FQH system. One can also perform the reverse, mapping the FCI
Hamiltonian such as a Hubbard type interaction $H_U$ of the
lattice fermions to a FQH Hamiltonian
$H_{\text{FQH}}=fH_Uf^{-1}$. More details of such a mapping will be evaluated in the following sections.

There is a subtle point that we want to discuss before ending
this section. The definition of maximally localized Wannier states in
Eq.~\ref{eq:MLWF} still leave an ambiguity in the relative phase
between different $\ket{W_{nk_y}}$. If we redefine
$\ket{W_{nk_y}}\rightarrow e^{i\theta_{k_y}}\ket{W_{nk_y}}$ with any
phase $\theta_{k_y}$, all the properties discussed above remain the
same. The map $f$, however, depends on this choice and different choices of phase corresponds to physically different mappings between FCI and FQH systems. To preserve the locality in the mapping, a choice should be made which makes $\ket{W_{nk_y}}$ continuous in $k_y$ in the large $L_y$ limit. An example of the choice is the following\cite{maissam}:
\begin{eqnarray}
\theta_{k_y}&=&-\sum_{0\leq p_y<k_y}A'_{p_y}-k_yP_{y0},\\
P_{y0}&=&-\frac1{2\pi}\sum_{0\leq p_y<2\pi}A'_{p_y},\nonumber\\
A'_{p_y}&=&-{\rm Im}\log\left(\sum_{\alpha}u_{0,p_y+\frac{2\pi}{L_y},\alpha}^*u_{0,p_y,\alpha}\right).\nonumber
\end{eqnarray}
In the $L_y\rightarrow \infty$ limit, $A'_{p_y}\simeq
a_y(0,p_y)\frac{2\pi}{L_y}$. This choice of $\theta_{k_y}$ corresponds
to a gauge transformation which makes $a_y({\bf k})$ uniform along the
$k_x=0$ line. Any other gauge choice
$\theta'_{k_y}=\theta_{k_y}+\delta\theta_{k_y}$ also works and
describes topologically equivalent states, as long as
$\delta\theta_{k_y}$ is a smooth periodic function of $k_y$ in the
large $L_y$ limit. While different gauge choices $\delta\theta_{k_y}$ of the Wannier states do not change the topological universality class of the associated state, it can be used as variational parameters in the many-body ground state and can be optimized numerically by the comparison of the Wannier ground state with the exact ground state.\cite{wu2012,gunnar}

\section{Pseudopotential representation of interactions}
\label{sec:pp}

In this section, we review the PP representation of
interactions in FQH systems, and then map it to the FCI using the map
established in~\eqref{mapFQHFCI}.
The PP approach was pioneered by Haldane in the
context of FQHE~\cite{haldane83prl605}:
Suppose we assume all
relevant degrees of freedom to be located in the LLL. When projected onto this subspace, which is a perfectly flat band, the kinetic energy is frozen out
and the Hamiltonian thus consists only of an interaction term reminiscent of
the Coulomb interaction
\begin{equation}
H=\sum_{i<j} V (\bs{r}_i-\bs{r}_j),
\end{equation}
where the sum extends over all pairs of particles. This means that the
interaction is fully characterized by its interaction energy between the
particles $i$ and $j$. The PPs are then defined by
\begin{equation}
H=\sum_{i<j} \sum_{m=0}^{\infty} V^m P^m_{ij},
\end{equation}
where $ P^m_{ij}$ projects onto a state where particles $i$ and $j$
have relative angular momentum $m$ and $V^m$ is the energy penalty
for having two particles in such a state, taking on odd values
$m=1,3, \dots 2\mathcal{M}+1$ for fermions and even values $m=0,2,\dots 2\mathcal{M}$ for bosons.   Since the LLL is perfectly flat, the $m$ index also characterizes the distance between the particles. As the interactions are of repulsive Coulomb
type, we can make the $V^m$s positive semidefinite by also fixing them to be real. This pseudopotential construction is particularly elegant
on the infinite plane and the sphere where it was first defined~\cite{haldane83prl605,fano-86prb2670},
where translational symmetry and rotational symmetry are preserved and the relative angular momentum of a two-electron state is thus a
well-defined quantum number. In particular, this construction allows for a most explicit
connection between the Hamiltonian and the clustering properties of
quantum Hall wave functions. For the Laughlin state wave
function at $\nu=1/3$ filling, where\footnote{$z_i$ represents
represents complex Cartesian coordinates in the case on a plane. On the sphere, $z_i\sim \frac{v_i}{u_i}$ where $u_i$ and $v_i$ are the spinor coordinates.}
$\ket{\Psi_{\text{L}}}\sim \prod_{i<j} (z_i-z_j)^3$, the wave function decays as the $3$rd
power as two particles approach each other. This shows that two
electrons are never allowed in a relative angular momentum state of $m=1$, i.e. the
particles maximally avoid each other within a featureless liquid
state~\cite{haldane83prl605}. It follows that $\ket{\Psi_{\text{L}}}$ is located in the
nullspace of $H_{\text{Laughlin}}=\sum_{i<j} V_1 P_{ij}^1$. Furthermore, the
Hilbert space at the appropriate filling only allows for one state
of such a kind on a trivial genus manifold, implying that $\ket{\Psi_{\text{L}}}$ is the
exact unique ground state of $H_{\text{L}}$. Similar trial Hamiltonian constructions are allowed
for many other states, including
non-Abelian quantum Hall states such as the Read-Rezayi series and others~\cite{Read-99prb8084,simon-07prb075318,mcs}.

The Landau level on a cylinder can be seen as a hybrid version
of the spherical and planar scenario. Starting from a Cartesian $(x,y)$
plane, we impose periodic boundary conditions along $y$ and, in analogy
to the sphere, quantize the pseudo-momentum along $y$ according to the total
magnetic flux $N_\phi$, constraining the available area for
the guiding center motion. We assume an infinite cylinder along the $x$-direction. On the cylinder geometry, the PP Hamiltonian takes
the explicit form~\cite{rezayi-94prb17199}
\begin{eqnarray}
H&=&\gamma \sum_{i<j}\sum_{n,m} \int  d q \; l_B V^m L_m(q^2l_B^2
+\gamma^2n^2) e^{-q^2 l_B^2/2} \nonumber \\
&& \times e^{-\gamma^2 n^2/2} e^{i\gamma n \hat{x}_i/l_B} e^{iq
  (\hat{y}_i-\hat{y}_j)} e^{-i\gamma n \hat{x}_j /l_B}, \label{cyl}
\end{eqnarray}
where $\gamma$ is the aspect ratio of the cylinder, $l_B$ is the
magnetic length, $q$ denotes the momentum variable along $x$, $V^m$ is the PP energy of a state with
relative angular momentum $m$, and $L_m$ denotes the $m$th Laguerre
polynomial~\cite{Abra}. (Note that we have explicity included all
units in \eqref{cyl} as compared to Ref.~\onlinecite{rezayi-94prb17199}.) Eq.~\ref{cyl}
gives a pair interaction energy for each
$m$. In the plane limit where $\gamma n \rightarrow k_xl$, the sum over
$n$ reduces to a momentum integral along $y$ analogous to the $q$ integration
along $x$, resulting in a two-dimensional momentum integral which reduces to the pair
interaction
\begin{eqnarray}
V(\bs{r}_i-\bs{r}_j)&=&\sum_m V^mP^m(\bs{r}_i-\bs{r}_j) \nonumber \\
&=& \sum_m V^m L_m (-l_B^2 \nabla^2) \delta^2 (\bs{r}_i -\bs{r}_j).
\label{lag}
\end{eqnarray}
This decomposition is valid for sufficiently short-ranged potentials $V(\bs{r_i}-\bs{r_j})$. Note that the functional form of Eq.~\ref{lag} is different
from that in Ref.~\onlinecite{rezayi-94prb17199} because we have used ordinary coordinates instead of guiding
center coordinates. If one replaces all
coordinates including their derivatives with their guiding center
analogues, $l_B^2 \delta^2 (\bs{r}_i -\bs{r}_j)$ will be replaced by the exponential tail expression
$e^{-\vert \bs{r}_i - \bs{r}_j \vert^2/2l_B^2}$ in Ref.~\onlinecite{rezayi-94prb17199}. The details of this
calculation are shown in Appendix~\ref{equi}.

For later purposes, we also Fourier transform Eq.~\ref{lag} and
invoke the orthogonality relation of the Laguerre polynomials (see also Appendix~\ref{nbody}) to obtain
\begin{equation}
V^m=4\pi l_B^2\int \frac{ d^2k}{(2\pi)^2} e^{-l_B^2 k^2}L_m(k^2l_B^2)V(k).
\label{Vm}
\end{equation}

The above expression, which will also be rederived in Appendix~\ref{nbody} as a special case of a much more general result obtained from first principles, enables us to compute the PP coefficients $V^m$ directly from a generic  potential $V(k)$. It is the starting point for the generalization to interactions involving more than two bodies, as is described in Section VI.

To apply the PP decomposition to an FCI system with periodic
boundary conditions in both directions, we have to compactify the open direction of
the cylinder. The single particle states on the torus are the
$\psi_K(x,y)$ defined in Eq.~\ref{torusLLWF}. In this basis, the $m$th PP Hamiltonian has the matrix elements
\begin{widetext}
\begin{equation}
U^m_{n_1,n_2,n_3,n_4}= \int d^2 \bs{r}_i d^2
\bs{r}_ j (\psi_{n_1}(\bs{r}_i)^* \psi_{n_2}(\bs{r}_j)^* -
\psi_{n_2}(\bs{r}_i)^* \psi_{n_1}(\bs{r}_j)^*) U^m(\bs{r}_i-\bs{r}_j) (\psi_{n_3}(\bs{r}_i) \psi_{n_4}(\bs{r}_j) -
\psi_{n_4}(\bs{r}_i) \psi_{n_3}(\bs{r}_j)).
\end{equation}
\end{widetext}
$U^m$ refers to a normalized PP that has nonzero
projection only in the $m$th relative angular momentum sector. They
form a basis in which a generic potential $V$ is expanded. As such,
the $V^m$s which appear in Eq.~\ref{Vm} and other places below refer
to the component of $V$ proportional to $U^m$, i.e. projected onto the
relative angular momentum sector $m$. For simplicity, we have denoted $\psi_{K=2\pi n/L_y}({\bf r})$ as
$\psi_n({\bf r})$, $n=1,2,...,L_xL_y$.

We can use the map
(\ref{mapFQHFCI}) defined in Sec. \ref{sec:wannier} to define the
corresponding PP Hamiltonian in FCI, which has the second-quantized form
\begin{eqnarray}
U^m&=&\sum_{n_1,n_2,n_3,n_4} a_{n_1}^{\dagger} a_{n_2}^{\dagger}
U^m_{n_1,n_2,n_3,n_4} a_{n_3}^{\phantom{\dagger}}
a_{n_4}^{\phantom{\dagger}}, \nonumber \\
&=&\sum_{n+l_1\in \mathbb{Z},n+l_2\in \mathbb{Z}} U_{l_1l_2}^{m} a_{n+l_1}^{\dagger} a_{n-l_1}^{\dagger}
 a_{n-l_2}^{\phantom{\dagger}}
a_{n+l_2}^{\phantom{\dagger}}. \label{1d}
\end{eqnarray}
Here,
\begin{eqnarray}
a_n=\sum_{i,\alpha}\braket{W_{2\pi n/L_y}}{i\alpha}c_{i\alpha}
\end{eqnarray}
 is the annihilation operator of the Wannier state $\ket{W_{2\pi n/L_y}}$. The matrix element $U^m_{n_1n_2n_3n_4}$ is simplified to the form of $U^m_{l_1l_2}=U^m_{n+l_1,n-l_1,n-l_2,n+l_2}$ due to the magnetic translation symmetry. (More discussion on the magnetic translation symmetry will be presented in Sec.~\ref{sec:CMMT}.) Depending on whether we consider the torus or cylinder geometry, the
sites along the main cylinder axis labelled by $n$ are assumed to obey
periodic or open boundary condition, respectively. For the cylindrical case, Eq.~\ref{1d} can be brought into a bosonic pair creation form given by
$U^m_{l_1 l_2}=g\kappa^3 b^m_{l_1}b^m_{l_2}$, where
\begin{equation}
b^{2j+1}_l = le^{-\kappa^{2}l^2}\sum_{p=0}^j \frac{(-2)^{3p-j}(\kappa l)^{2p}\sqrt{(2j+1)!}}{(j-p)!(2p+1)!}
\end{equation}
so that
\begin{equation}
U^m=g\kappa^3\sum_n \hat{b}^{m\dagger}_n \hat{b}^m_n,
\label{PP2}\end{equation}
where $\hat{b}_n^m=\sum_{n+l \in \mathbb{Z}} b ^m_{l}a_{n-l}a_{n+l}$. Here,
$g=\frac{4V_0}{(2\pi)^{3/2}}$ is a constant with units of energy and
$\kappa=\frac{2\pi l_B}{L_y}=\frac{1}{L_y}$. $l_B$ has been set to
$\frac{1}{2\pi}$ lattice constants in the latter equality in accordance to
Ref.~\onlinecite{qi11prl126803}. In the following, $l_B$ will be expressed in
units of the lattice constant unless it appears in the combination
$l_B^2 \nabla^2$ or $l_B^2 q^2$, where $q$ is a momentum variable. The complete derivation of Eq.~\ref{PP2} can be found in
Appendix~\ref{2body}.

The Hamiltonian in~\eqref{1d} will be the
starting point of Sec.~\ref{sec:model} when we expand different FCI
models into this PP form. Its $m=1$ case has been previously used to define low-dimensional Mott-type models with bare onsite hardcore potentials at fractional filling~\cite{lee-04prl096401,seidel-05prl266405}.
The PP $U^m_{\text{tor}}$ on the torus can be found by summing
over all the periodic images of $U^m_{l_1l_2}$ (referred to as
$U^m_{\text{cyl}}$ in Appendix~\ref{ortho}) satisfying $l_1 +
l_2$ $\text{mod}$ $2L_xL_y=0$. This constraint can be generalized to the case with more
than two bodies, as shown in Appendix~\ref{3body}.

For
finite-size investigations on the cylinder or the torus, we have to
keep in mind that relative angular momentum is no longer a
well-defined quantum number, as opposed to the case of the sphere or the
plane.
The parameter $m$ in~\eqref{lag}, which corresponds
to the exact relative angular momentum as we take the planar limit,
determines the order of
the derivative acting on the hardcore potential via the degree of the Laguerre polynomial. This corresponds to a
Taylor expansion of the interaction in momentum
space~\cite{trugman-85prb5280}. While its interpretation as the
exact relative angular momentum is absent, it can still be employed as
an expansion parameter for short-range interactions on a sufficiently
large torus or cylinder.
To see this in terms of the Hilbert space basis, we describe
relative motion states on the torus by relative motion states on the
plane. The latter can be exactly classified via the relative angular
momentum $m$ which is
proportional to the interparticle distance in that relative state
$r_m$.
For $r_m/L_x, r_m/L_y << 1$, the overlap of the torus and planar
relative motion states goes to unity, effectively
reestablishing the notion of torus relative angular momentum for short distances.
Still, this approximation
becomes invalid for higher values of relative
angular momentum. At the Hamiltonian level, this is reflected by the overcompleteness of the PPs $U^m$.
This occurs because the interparticle distance $r_m\sim L_x,L_y$ that characterizes a relative angular momentum state
will no longer be well-defined when $r_m$ is comparable to the system lengths of the torus. A quantitative treatment of the
overcompleteness bounds can be found in Appendix~\ref{ortho}.

A deep insight to note is that even though an exact
angular momentum quantum number cannot be defined, the clustering
property of the quantum Hall-type wave functions are still fixed
appropriately at these finite size manifolds such that they can be
{\it exactly} located in nullspaces of PP
Hamiltonians. This was
elegantly shown for the torus by Haldane and
Rezayi~\cite{haldane-85prb2529}, which we illustrate for the $U_1$ PP at $\nu=1/3$ filling: demanding that the many-particle
ground state pays no energy due to $U_1$, it necessitates that the
wave function decays to third power as the particles approach each
other. Oddness due to fermionic statistics and boundary conditions on
the torus automatically restricts the functional form of the wavefunction to be $\Psi \sim
\prod_{i<j} \vartheta(z_i-z_j\vert \tau)^3 $, where $\vartheta$ is the odd Jacobi
theta function. The groundstate is thus forced to be of Laughlin type. This fixes $N_\phi-3$ zeroes of the wave function,
where the remainder $3$ constitute the topological center of mass
degeneracy of the Laughlin state~\cite{wen-09prb9377}. A similar discussion can be pursued
for the cylinder, where the center of mass degeneracy is absent but
the clustering property of the wave function leads to the same finding
for the remainder functional form of the wave function~\cite{bergholtz-08prb155308}.

All in all, the properties of the pseudopotential expansion sets the stage for numerical investigations of short-ranged
interactions of FCIs as well as the defining of trial Hamiltonians for new quantum
Hall-type fractional Chern phases, both of which will be pursued in the following.

\section{Pseudopotential expansion of fractional Chern insulators}
\label{sec:model}

\subsection{Model Hamiltonians}

We apply the PP expansion to two model FCI Hamiltonians, the checkerboard (CB) model introduced in Ref.~\onlinecite{sun-11prl236803} and the honeycomb (HC) model introduced in Refs.~\onlinecite{haldane88prl2015,shengboson2011}.  Both models possess an almost flat (dispersionless) band which mimics the LLL in an FQH system. There, the Coulomb-type electron interactions lift the macroscopic degeneracy of the LLL, leading to a topologically degenerate groundstate. In the same spirit, we add Hubbard-type interaction terms to our model Hamiltonians such that
\begin{eqnarray}
H_{\text{int}}&=&\lambda H_{NN}+(1-\lambda)H_{NNN} \nonumber \\
&=&h_0 \left(\lambda \sum_{\langle ij \rangle}n_in_j +
  (1-\lambda)\sum_{\langle \langle ij
    \rangle\rangle}n_in_j\right),\nonumber \\
H&=&H_{0}+H_{\text{int}},
\label{hint_nonint}
\end{eqnarray}
where $\lambda$ characterizes the relative strengths of the nearest
neighbor (NN) and next-nearest-neighbor (NNN) interaction
terms. $h_0$, a parameter with units of energy, sets the overall
magnitude of $H_{\text{int}}$. The single-particle term $H_{0}$ gives
rise to the almost flat band and provides information for the
construction of the Wannier basis. This is the basis we will use for
expressing the two-body interacting term $H_{\text{int}}$ in the same
basis as the PP Hamiltonians $U^1, U^3,$ etc. as denoted
in~\eqref{1d}.

We  consider interactions that are much larger than the bandwidth
of the almost flat band, but much smaller than the interband gap. In
this limit, for a partially filled flat band, we can ignore the
coupling to the upper band and only study the effect of interactions
in the subspace of the flat band. With this picture in mind, we expand
$H_{\text{int}}$ in terms of the PPs in the Wannier basis of the
partially filled band. As long as the bandwidth of the almost flat band is much smaller than the interaction strength, we can ignore the bandwidth and consider only the interaction term.

\begin{figure}
\begin{minipage}{0.99\linewidth}
\includegraphics[width=0.35\linewidth]{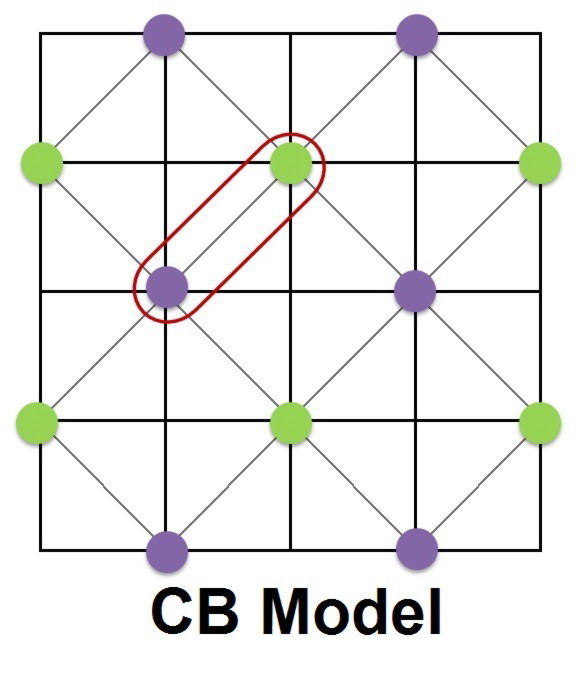}
\includegraphics[width=0.55\linewidth]{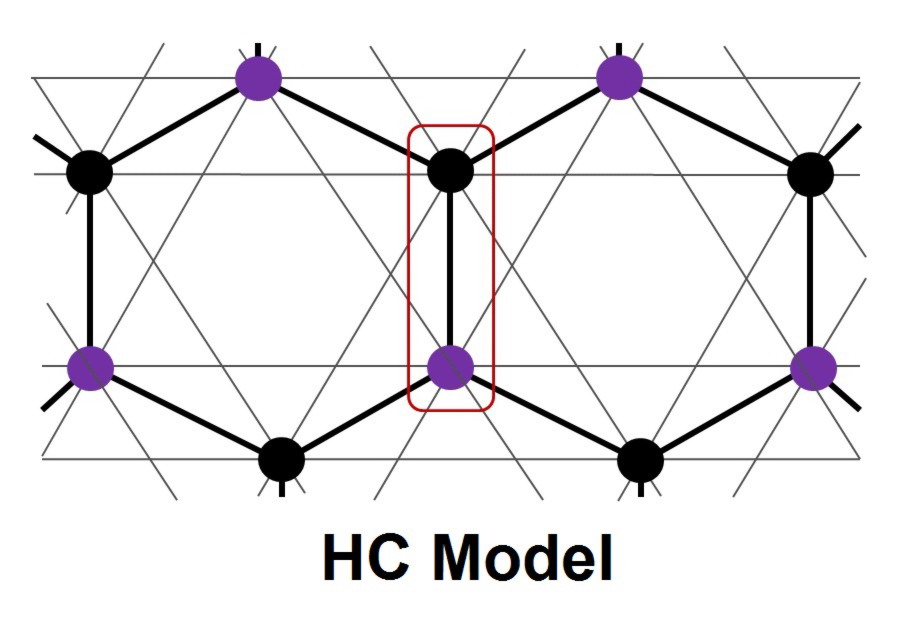}
\end{minipage}
\caption{(Color online) Lattice structure of the CB(left)  and HC (right)
  models. Sites colored differently belong to different
  sublattices. The unit cells are demarcated in red. For the CB model,
  NN interactions are between different sublattices while NNN and NNNN
  interactions occur within the same sublattice. For the HC model,
  both the NN and NNNN interactions occur between different
  sublattices, but the NNN interactions act within the same
  sublattice.}
\label{fig:CBHC}
\end{figure}

The checkerboard (CB) lattice model consists of two interlocking
square lattices displaced $(1/2,-1/2)$ sites relative to each
other (Fig.~\ref{fig:CBHC}). Its noninteracting Hamiltonian $H_0^{\text{CB}}$ consists of
NN, NNN and NNNN hopping terms parametrized by hopping
strengths $t$, $t'$, and $t''$ respectively~\cite{remarksun}. The NN hoppings exist between sites belonging to different sublattices and carry a phase $\phi$, giving rise to the time-reversal symmetry breaking necessary for a nonzero Chern number. Both the NN and NNNN hoppings exist between different sublattices, leading to off-diagonal terms in the single-particle (noninteracting) Hamiltonian. In sublattice space,
\begin{equation}
H^{\text{CB}}_{0}(k)=d_0 I + \sum_i d_i \sigma_i,
\label{modelhcb}
\end{equation}
where
\[d_1=-4t \cos \phi \cos \frac{k_x}{2} \cos \frac{k_y}{2},\]
\[d_2=-4t\sin \phi \sin\frac{k_x}{2}\sin \frac{k_y}{2},\]
\[d_3=-2t'(\cos k_x-\cos k_y).\]
 The expression for $d_0$ is irrelevant because it is not needed for
 the computation of the Wannier basis. We set $t=1$,
 $t'=-t''=1/(2+\sqrt{2})$ and $\phi=\pi/4$ as in
 Ref. \onlinecite{sun-11prl236803} to achieve the maximal the flatness ratio of $\sim 30$ for the
 bottom band. We can explicitly see why a nonzero $\phi$ is necessary
 for having a topologically nontrivial model: as the Chern
 number is given by $C_1=\frac{1}{4\pi}\int d^2k \vec \hat d \cdot
 (\partial_x\vec \hat d \times \partial_y\vec\hat d)$, it can only be nonzero if none of the $d_i$'s is identically zero.

Notice that $H^{\text{CB}}_{0}$ is not of Bloch form since the $d_i$'s
do not obey the periodicity of $2\pi$. This is because some sites are noninteger
lattice spacings away from each other (Fig.~\ref{fig:CBHC}). We can remedy this by shifting one sublattice site on top of the other within a unit cell. Mathematically, this corresponds to a gauge transformation of $c_{kB}^\dagger\rightarrow c_{kB}^\dagger e^{-i(k_x-k_y)/2}$ where $B$ refers to one of the sites within the sublattice. After the gauge transformation,
\[ d_1= -t[\cos\phi + \cos(k_x+k_y+\phi)+\cos(k_x-\phi)+\cos(k_y-\phi)], \]
\[ d_2= -t[\sin\phi + \sin(k_x+k_y+\phi)+\sin(k_x-\phi)+\sin(k_y-\phi)], \]
\[ d_3= -2t'(\cos k_x-\cos k_y).\]
The noninteracting part of the honeycomb model is defined
similarly. The unit cell consists of two adjacent sites. The phase
$\phi$ is carried between NNN sites, which lie in the same
sublattice. NNNN interactions which occur for diametral sites on the
same hexagon involve different sublattices (Fig.~\ref{fig:CBHC}). After an analogous gauge transformation,
\begin{equation}
H^{\text{HC}}_{0}(k)=d_0 I + \sum_i d_i \sigma_i,
\label{modelhhc}
\end{equation}
where
\[ d_1= -t(1+\cos k_x+\cos k_y)-t''(\cos(k_x+k_y)+2\cos(k_x-k_y)), \]
\[ d_2 = t(\sin k_x + \sin k_y) + t''\sin(k_x+k_y), \]
\[ d_3 = 2t'\sin\phi(\sin k_y-\sin k_x +\sin(k_x-k_y)). \]
The values for the
NN, NNN, and NNNN hoppings are given by $t=1$, $t'=0.6$, and
$t''=-0.58$, $\phi=0.4 \pi$ such that
the flatness ratio of the band is optimized to about $60$~\cite{shengboson2011}. We stress that while the optimization of these
flatband parameters is not necessary for performing the
PP expansion, it is physically relevant in
increasing the stability of an FQH state present in the system.

\subsection{Expressing $H_{\text{int}}$ in the Wannier basis}

We now have the necessary ingredients for expressing the interaction part of our model Hamiltonians in the same basis as the PPs in FQHE. First, we can perform a Fourier transform on $H_{\text{int}}$ such that it becomes
\begin{eqnarray}
H_{\text{int}}&=&\lambda h_0\sum_{\langle ij \rangle}n_in_j + (1-\lambda)h_0\sum_{\langle \langle ij \rangle\rangle}n_in_j \nonumber\\
&=& \sum_qV^{\alpha\beta}(q)n_{-q\alpha}n_{q\beta} ,
\end{eqnarray}
where $q$ is an internal momentum variable, $\alpha,\beta$ are the sublattice
indices, and $n_{q\alpha} =
\sum_{k_xk_y}c^\dagger_{k+q,\alpha}c_{k\alpha}^{\phantom{\dagger}}$. $V^{\alpha \beta}(q)$ denotes the $q$th
Fourier component of the interaction between the sublattice index $\alpha$
and $\beta$. This is an expression quartic in the creation and annihilation operators $c,c^\dagger$ in the momentum/sublattice basis. Since we are only considering interactions within the flat band, we project out the upper band and keep only the matrix elements of $n_{q\alpha}$ in the flat band. After projecting out the upper band, the annihilation operator $c_{k\alpha}$ can be expanded in the Wannier state basis:
\begin{eqnarray}
c_{\vec{k}\alpha}&=&\sum_{n}\braket{\vec{k}\alpha}{W_{2\pi n/L_y}}a_n +\text{upper band contributions}\nonumber\\
&\rightarrow &\sum_{n}\braket{\vec{k}\alpha}{W_{2\pi n/L_y}}a_n \nonumber\\
&=:&\sum_{n} U_{xk_x\alpha k_y} a_{x,k_y},
\label{normbase}
\end{eqnarray}
since $\frac{2\pi n}{L_y}=K=k_y+  2\pi x C_1=k_y+  2\pi x $. In this representation, the density operator becomes
\begin{eqnarray}
n_{q\alpha} &=& \sum_{k_xk_y}L_x^{-1/2}c^\dagger_{k+q,\alpha}c_{k\alpha} \nonumber \\
&=&\sum_{k_xk_y}L_x^{-1/2}\sum_{x_1}U^*_{x_1,k_x+q_x,\alpha,k_y+q_y}a^\dagger_{x_1,k_y+q_y}\nonumber\\
&&\sum_{x_2}U_{x_2k_x\alpha k_y}a_{x_2k_y}\nonumber \\
&=&\sum_{k_y x_1x_2}N_{q\alpha x_1x_2k_y}a^\dagger_{x_1,k_y+q_y}a_{x_2k_y}^{\phantom{\dagger}},
\nonumber \\
\end{eqnarray}
where the normalization factors $N_{q\alpha x_1x_2k_y}$ follow from~\eqref{normbase}.
We obtain
\begin{widetext}
\begin{eqnarray}
H_{\text{int}}&=&\sum_{q\alpha\beta}V^{\alpha \beta}_qn_{-q,\alpha}n_{q\beta} \nonumber \\
&=&\sum_{q_x}\sum_{\{x_j\},\{K_{jy}\},\alpha\beta}V^{\alpha\beta}_q\delta_{q_y-k_{2y}+k_{1y}}\delta_{k_{2y}-k_{1y}-k_{3y}+k_{4y}}N_{-q\alpha x_1 x_2k_{2y}}N_{q\beta x_3x_4k_{4y}}
a^\dagger_{K_1} a_{K_2}^{\phantom{\dagger}} a^\dagger_{K_3} a_{K_4}^{\phantom{\dagger}} \nonumber \\
&=&\sum_{K_1K_2K_3K_4}(h^{int})_{K_1K_2K_3K_4}a^\dagger_{K_1}
a_{K_2}^{\phantom{\dagger}} a^\dagger_{K_3} a_{K_4}^{\phantom{\dagger}} \nonumber \\
&\simeq& -\sum_{n'n}\sum_{l_1l_2}f(n,n',l_1,l_2)(a_{(n-l_1)/2}a_{(n+l_1)/2})^\dagger(a_{(n'-l_2)/2}a_{(n'+l_2)/2}).
\label{basischange}
\end{eqnarray}
\end{widetext}
A quadratic term has been dropped in the final step because it
can be absorbed into the noninteracting part of the Hamiltonian. The latter is irrelevant for our current purpose of expressing the interaction operator in the Wannier basis. Note, however, that this quadratic term should not be omitted if we were to perform studies on energetics. Due to fermionic statistics of the $c_{n\pm l}$ operators, we can antisymmetrize $H_{\text{int}}$, leading to the matrix elements
\begin{eqnarray}
h(n,n',l_1,l_2)&=&f(n,n',l_1,l_2)-f(n,n',-l_1,l_2) \nonumber \\
&&-f(n,n',l_1,-l_2)+f(n,n',-l_1,-l_2),\nonumber \\
\end{eqnarray}
which are manifestly antisymmetric in $l_1$ and $l_2$, just like the
PPs $U^m_{l_1l_2}$. We see from Eq.~\ref{basischange}
that $h(n,n',l_1,l_2)$ corresponds to a pair hopping interaction on a
line. Two particles with the CM ``position" $n'$
separated by $l_2$ sites simultaneously hop onto new positions with CM
position $n$ separated by $l_1$ sites (see also Fig.~\ref{fig:nonconservation}). More discussions on the physical interpretation of this interaction will be presented in Section~\ref{sec:CMMT}.

\subsection{Pseudopotential expansion of the interaction term}

While the PP matrix elements $U^m_{l_1l_2}$ depend only
on $l_1$ and $l_2$, the FCI interaction Hamiltonian matrix elements in
the Wannier basis $h(n_1,n_2,l_1,l_2)$ also depend on $n_1$ and
$n_2$. As a consequence, only a part of $h(n_1,n_2,l_1,l_2)$ can be expanded in terms of PPs. This important fact can be understood in terms of magnetic translation (MT) symmetry breaking, which will be analysed in depth in the next section. Here, we shall concern ourselves with the terms that can be expanded in PPs, defined by
\begin{equation}
H_p(n,n',l_1,l_2)=\frac{\delta_{n n'}}{2L_xL_y}\sum_{N=1}^{2L_y} h(N,N,l_1,l_2).
\label{hp}
\end{equation}
$H_p$ vanishes for $n \neq n'$ and does not depend on $n$ when $n_1=n_2=n$, as required. The sum runs from $1$ to $2L_y$ because $h(n,n,l_1,l_2)$ is periodic in $n$ with period $2L_xL_y$, as evident from the periodicity of $a_{(n\pm l)/2}$ in Eq.~\ref{basischange}.

We would like to expand $H_p$ in an orthonormal basis of
PPs $U^1$, $U^3$, etc. However, this expansion is only
unique and thus meaningful if we include PPs with $m$ bounded by a certain $m_{\rm max}$. This is because the inclusion of
higher PPs can yield an overcomplete operator basis, a
consequence of the finite size of the torus geometry explained in Section~\ref{sec:pp}. The truncated PP basis is no longer complete, but we can still perform a PP expansion of $H_p$ (now suitable normalized) by writing
\begin{equation}
H_p=\sum^{m_{max}}_{m} V^m U^m+H_{>} =:H_{\text{pseudo}} +H_{>},
\end{equation}
and finding the PP expansion coefficients $V^m$ that maximize the normalized
overlap $\langle H_p , H_{\text{pseudo}} \rangle$. The overlap is taken by
summing over all $|l_1|,|l_2|\leq L_xL_y$ since $H_p$ has a period of
$2L_xL_y$. Specifically, for any two Hamiltonians $H$ and $H'$ that respect MT symmetry,
\begin{equation}
\langle H,H'\rangle =
\frac{\sum_{l_1l_2}H_{l_1l_2}H'_{l_1l_2}}{\sqrt{\sum_{l_1l_2}H^2_{l_1l_2}\sum_{l_1l_2}H'^2_{l_1l_2}}}.
\label{over}
\end{equation}
The term $H_{>}$ consists of the part of $H_{\text{int}}$ that
does not break MT symmetry,  but still cannot be uniquely expressed in
terms of the PPs. It includes, for instance, hoppings
that occur over lengths comparable to the size of the torus.

When the $U^{m}$'s form an orthonormal basis, the $V^m$'s that maximize the normalized overlap $\langle H_p,H_{\text{pseudo}}\rangle $ can be determined as
\[V^m=\langle H_{p},U^{m}\rangle .\]

\begin{figure}
\begin{minipage}{0.99\linewidth}
\includegraphics[width=\linewidth]{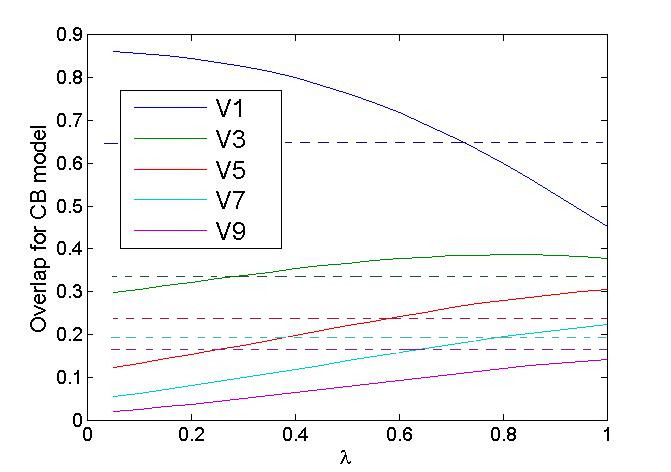}
\includegraphics[width=\linewidth]{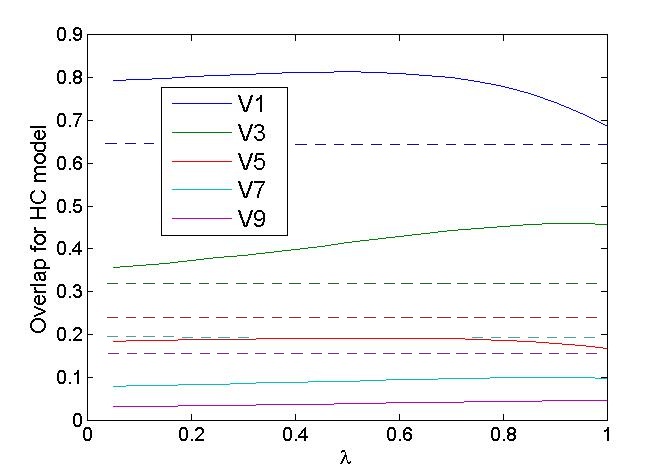}
\end{minipage}
\caption{(Color online) The pseudopotential expansion of the fermionic checkerboard (upper) and the
  honeycomb (lower) models for $L_x=L_y=6$. The normalized overlap $V^j$ is
  plotted against $\lambda$, where $H_{\text{int}}=\lambda H_{NN} +
  (1-\lambda) H_{NNN}$, so that we have the NNN limit on the left and the NN limit on the right.
  For the CB model, we see that the NN and NNN terms exhibit marked diferences in their pseudopotential expansions,
  with the MT symmetry conserving part of its NNN term consisting
  almost exclusively of the $V^1$ and $V^3$ terms. This can be
  understood by studying its distribution of matrix elements (Fig.~\ref{fig:l1l2plots}). For comparison, the PP coefficients $V^1$ to $V^9$ of the QH Coulomb interaction are plotted as dashed lines. The NNN interaction has a larger $V^1$ coefficient and smaller $V^3,V^5,...$ coefficients than the Coulomb interaction, and is hence even more likely to exhibit Laughlin groundstate.
 }
\label{fig:V13579}
\end{figure}
\begin{figure}
\begin{minipage}{0.99\linewidth}
\includegraphics[width=.28\linewidth]{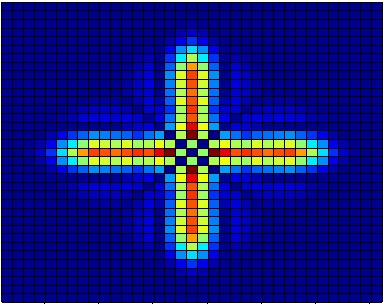}
\includegraphics[width=.28\linewidth]{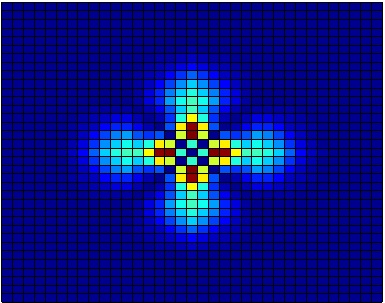}
\includegraphics[width=.28\linewidth]{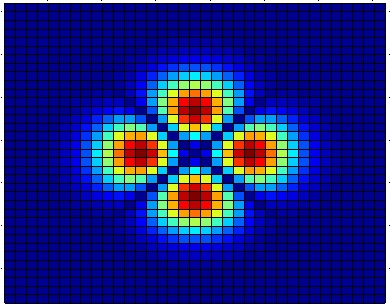}
\includegraphics[width=.1\linewidth]{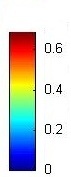}
\end{minipage}
\caption{(Color online) Plots of normalized $|H_p|$ for the NN (left) and NNN (center)
  terms of  for the fermionic CB model. The horizontal and vertical axes
  represent $l_1-l_2$ and $l_1+l_2$ respectively. Regions with
  relatively large $|H_P|$ are colored red. $U^1$ (right) is plotted
  for comparison. The NNN term evidently bears more resemblance to
  $U^1$. }
\label{fig:l1l2plots}
\end{figure}

From Fig.~\ref{fig:V13579}, we see that the percentage of $H_p$ that
can be expanded as PPs
$\sum_{j=0}^{j_{max}}\left|V^j\right|^2$ has the maximum value for
$0.94$ for the CB NNN interaction (Fig.~\ref{fig:V13579}). As
expected, the first PP has the highest weight, which favors the possibility of simple FQH states such as Laughlin states. With the relative angular momentum $m$ being proportional to interparticle distance, $U^m$ is expected to
decay faster with $|l_1|,|l_2|$ as $m$ increases. This will be
shown in more detail in Appendix~\ref{2body}. It is notable, however, that the
second neighbor coupling leads to a better overlap with the first
PP than the nearest neighbor Hamiltonian. This
indicates that the PP Hamiltonians mapped to FCI systems
are not simple density-density interactions and their matrix elements
in real space lattice site basis can exhibit a nonmonotonic dependence
on distance. More specifically, this is because $U^1_{l_1l_2}\sim l_1l_2$ is not strongly peaked around  $l_1=\pm l_2$ in $l_1-l_2$ space, as shown in Fig.~\ref{fig:l1l2plots}, unlike the NN interaction. Since $l_1=\pm l_2$ corresponds to $q=0$ (as defined in Eq.~\ref{basischange}), we see that the NN terms are "too local" for a good overlap with $U^1$.
In general, the matrix elements $U^m_{l_1l_2}\sim (l_1l_2)^m$, so
$U^m$ becomes more localized at $l_1=\pm l_2$ for higher $m$.

For comparison, the pseudopotential coefficients for the Coulomb interaction in a QH system are also plotted in Fig.~\ref{fig:V13579}. They can be derived via Eq.~\ref{Vm}, where $V(k)=\frac{4\pi}{k}$. We see that the PP coefficients of the FCI interactions do not differ too much from those of the Coulomb interaction, and in fact have a larger $V^1$ coefficient in a large range of $\lambda$.

\subsection{Fermion-Boson Asymmetry}
In the quantum Hall effect, a Vandermode determinant allows to
equivalently switch from bosons to fermions which corresponds to
an additional attachment of one flux per particle. This symmetry is
broken in the fractional Chern insulator. We can see this explicitly by comparing the
PP coefficients of both the fermionic and bosonic HC model. The latter model is also studied in
other works like Ref. \onlinecite{shengboson2011}. The bosonic PPs are constructed analogously to the fermionic ones,
except that they are now symmetrized instead of antisymmetrized (refer to Appendix C for more details).

The comparison between the PPs of the bosonic and fermionic HC models are displayed in Fig. 5. The bosonic PP coefficients are in general closer to each other, with $V^0$ not larger than $V^2$. This is because of the large MT symmetry breaking (further described in the next section) that renders even the NN term rather nonlocal in the $l_1$, $l_2$ basis.

\begin{figure}
\begin{minipage}{0.99\linewidth}
\includegraphics[width=.97\linewidth]{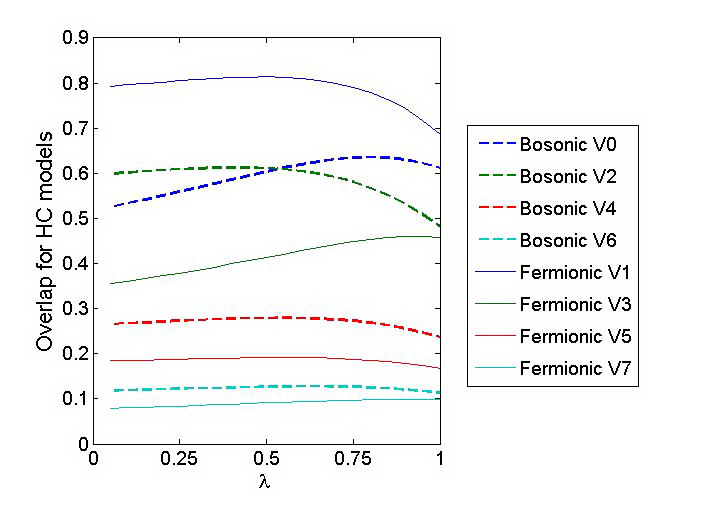}

\end{minipage}
\caption{(Color online) The pseudopotential expansion of the fermionic (solid line) and bosonic (dashed line) HC models for $L_x=L_y=6$.
The PP coefficient $V^j$ is
  plotted against $\lambda$, where $H_{\text{int}}=\lambda H_{NN} +
  (1-\lambda) H_{NNN}$, so that we have the NNN limit on the left and the NN limit on the right. The bosonic PP coefficients are in general closer to each other, with $V^0$ not larger than $V^2$ for some values of $\lambda$.}
  \end{figure}

\section{The effect of Magnetic Translation Symmetry breaking}
\label{sec:CMMT}

In this section, we analyze the origin of the terms in the FCI
Hamiltonian that cannot be expanded into PPs. We review how magnetic translation (MT) symmetry in FQH system
constrains the form of its two-body interaction terms, and investigate
how this picture is generalized to the FCI case.

\subsection{Origin of Magnetic Translation Symmetry breaking}

Consider an $L_xl_B\times L_yl_B$ torus geometry which has been discussed in Sec. II. In the Landau gauge $A_x=0,~A_y=Bx$, the covariant momentum operators are $P_x=-i\partial_x,~P_y=-i\partial_y-A_y$ which satisifes $\left[P_x,P_y\right]=iB$. The Hamiltonian $H=\frac1{2m}\left(P_x^2+P_y^2\right)$ has two translation symmetries $T_x^B,T_y^B$ defined by
\begin{eqnarray}
T_x^B&=&e^{i\frac{2\pi y}{L_yl_B}}e^{iP_x\frac{l_B}{L_y}},\nonumber \\
T_y^B&=&e^{iP_y\frac{l_B}{L_x}}.
\end{eqnarray}
$T_y^B$ is an ordinary translation while $T_x^B$ is a translation in $x$ direction by $l_B/L_y$ accompanied by a gauge transformation. The translation can only be defined in units of $l_B/L_y$ so that the change of gauge potential $A_y=Bx\rightarrow B(x+l_B/L_y)$ can be cancelled by a gauge transformation. The action of $T_x^B,T_y^B$ on the basis wavefunctions (\ref{torusLLWF}) is
\begin{eqnarray}
T_y^B\ket{\psi_K}&=&e^{iK/L_x}\ket{\psi_K},\nonumber \\
T_x^B\ket{\psi_K}&=&\ket{\psi_{K+\frac{2\pi}{L_y}}}.\label{magtrans}
\end{eqnarray}
For a general two-body interaction with $H_{\rm int}$ in the form of
$H_{\rm int}=\sum_{n_1,n_2,n_3,n_4}a_{n_1}^\dagger a_{n_2}^\dagger
U^{n_1n_2n_3n_4}a_{n_3}a_{n_4}$,  the condition $\left[T_y^B,H_{\rm
    int}\right]=0$ requires $n_1+n_2=n_3+n_4$, since
${(T_y^B)}^{-1}a_{n_1}^\dagger a_{n_2}^\dagger
a_{n_3}a_{n_4}{T_y^B}=a_{n_1}^\dagger a_{n_2}^\dagger
a_{n_3}a_{n_4}e^{2\pi i(n_1+n_2-n_3-n_4)/L_xL_y}$. The condition
$\left[T_x^B,H_{\rm int}\right]=0$ requires
$U^{n_1n_2n_3n_4}=U^{n_1+1,n_2+1,n_3+1,n_4+1}$. Therefore, the magnetic
translation symmetry $T_x^B$ and $T_y^B$ determines the
CM conservation ($n_1+n_2=n_3+n_4$ or $n=n'$) and (one-dimensional)
translation symmetry of the interaction Hamiltonian in FQH states,
i.e., $n$-independence of the interaction matrix elements.

\begin{figure}
\begin{minipage}{0.99\linewidth}
\includegraphics[width=.75\linewidth]{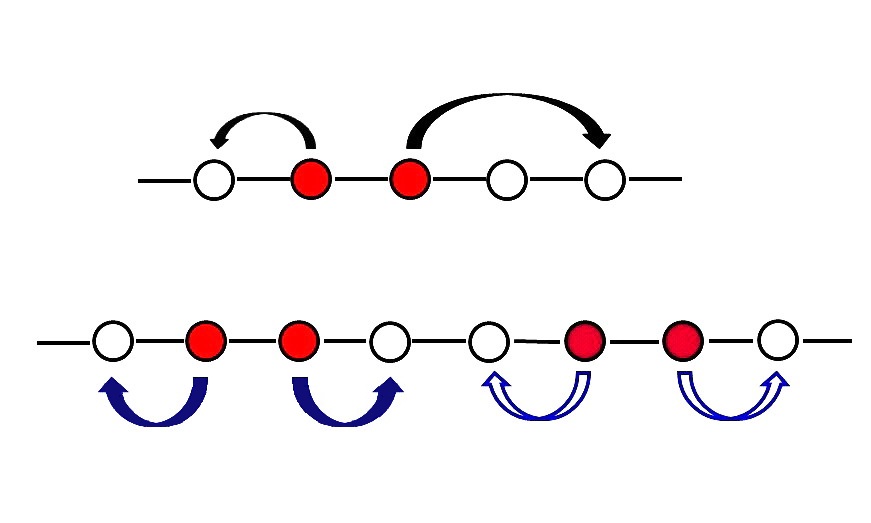}
\end{minipage}
\caption{(Color online) The two types of MT breaking hoppings. Top: A hopping process that changes the CM position. Bottom: Two hopping processes that preserve the CM position but break magnetic translation symmetry because the hopping at different CM position $n$ (solid and hollow arrows) have different amplitude.
}
\label{fig:nonconservation}
\end{figure}

By comparison, in the lattice model, we only have the lattice translation symmetries which commute with each other. The action of the lattice translation $T_x,T_y$ acts on the Wannier basis as
\begin{eqnarray}
T_y\ket{W_K}&=&e^{iK}\ket{W_K},\nonumber \\
T_x\ket{W_K}&=&\ket{W_{K+2\pi}}.\label{latticetrans}
\end{eqnarray}
Comparing Eq.~\ref{latticetrans} with Eq.~\ref{magtrans}, we see that in the mapping from FCI to FQH defined in Sec.~\ref{sec:wannier}, $T_x, T_y$ is mapped to $\left(T_x^B\right)^{L_y}$ and $\left(T_y^B\right)^{L_x}$, respectively. Therefore, in the lattice model, the translation symmetries only require the matrix element of two-body interaction $U^{n_1n_2n_3n_4}$ to satisfy
\begin{eqnarray}
U^{n_1n_2n_3n_4}&=&U^{n_1+L_y,n_2+L_y,n_3+L_y,n_4+L_y},\nonumber \\
U^{n_1n_2n_3n_4}&=&0\text{~if~}n_1+n_2\neq n_3+n_4\text{~mod~}L_y.
\end{eqnarray}
The magnetic translation symmetry breaking in the lattice models (Fig.~\ref{fig:nonconservation}) is
also related to the non-uniform Berry curvature in momentum
space. As was discussed in Sec.~\ref{sec:wannier}, the CM
position of the Wannier state $\ket{W_K}$ is determined by the flux of
the Berry's phase gauge field $P_x(k_y)$. If the system has magnetic
translation symmetry, $\ket{W_K}$ and $\ket{W_{K+2\pi/L_y}}$ are
related by $T_x^B$, so that $P_x(k_y)$ must depend on $k_y$
linearly. As a result, we expect MT symmetry breaking whenever the Berry curvature is nonuniform in momentum space, which is the case in a generic CI.

In addition, MT symmetry breaking will still be present even in the hypothetical case of perfectly flat Berry curvature. This is because the Wannier basis is not perfectly local. Recall from~\eqref{basischange} that
\begin{equation} n_1+n_2=n{\phantom{'}}=k_{1y}+k_{2y}+(x_1+x_2)L_y, \label{eqn1}\end{equation}
\begin{equation} n_3+n_4=n'=k_{1y}+k_{2y} + (x_3 + x_4)L_y, \label{eqn2}\end{equation}
where the $x_i$s are the lattice sites of the original
$H_{\text{int}}$. CM nonconserving terms occur where $x_1+x_2 \neq x_3
+ x_4$, when $n$ and $n'$ differ by a multiple of $L_y$. These terms
do not appear in the original real-space basis where
$H_{\text{int}}\propto \sum_{ij}n_in_j = -\sum_{ij}c^\dagger_i
c^\dagger_j c_i c_j$ annihilates and creates two particles at the same
position. However, our Wannier basis functions generically have
exponentially decaying tails on both sides of their peak $\bar x$, which produce CM nonconserving and thus MT breaking contributions.

\subsection{Numerical results on MT symmetry breaking}

\begin{figure}
\begin{minipage}{.99\linewidth}
\includegraphics[width=.9\linewidth]{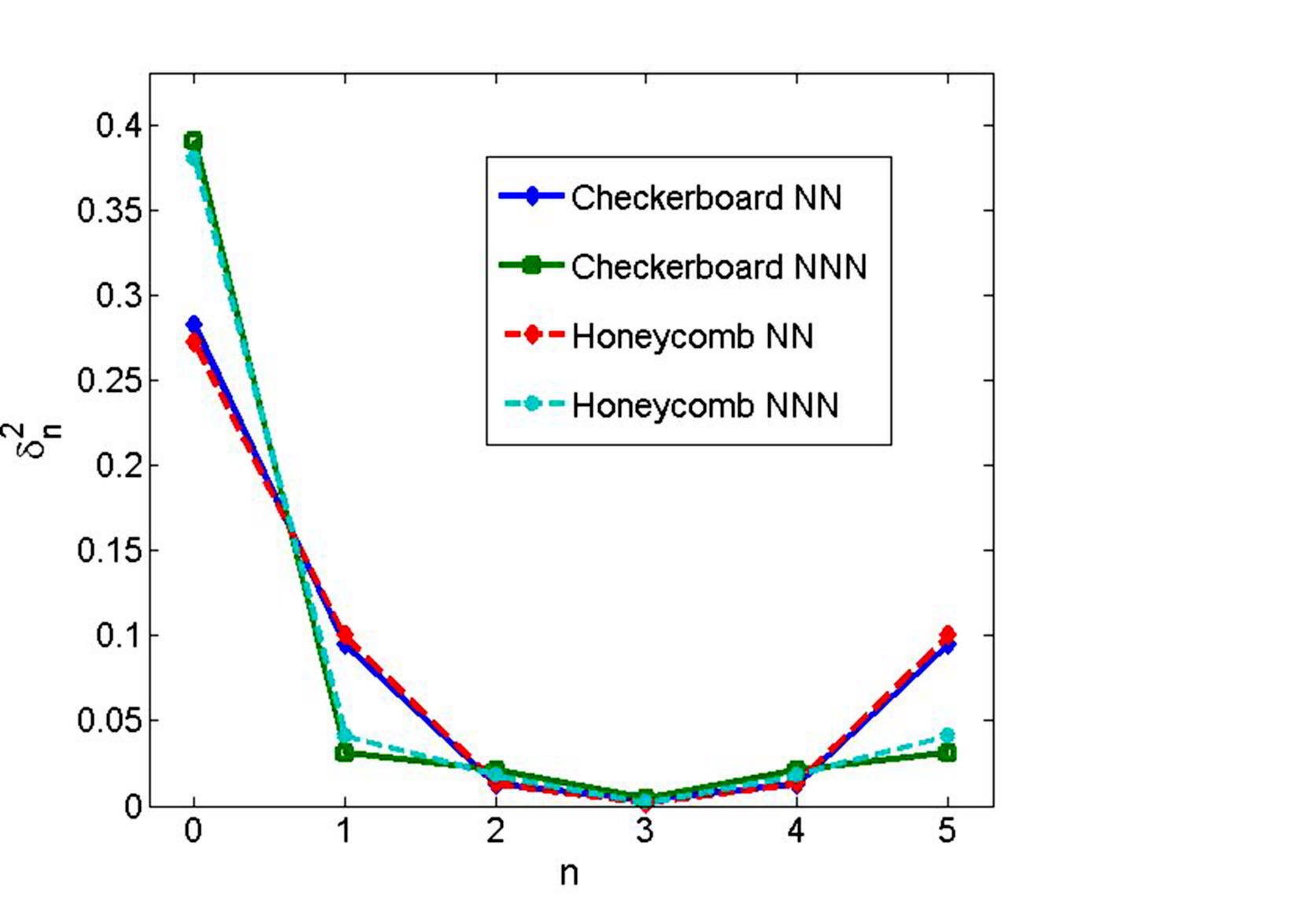}
\end{minipage}
\caption{(Color online) A plot of $\delta_n^2$ for $L_x=L_y=6$ for
  different model Hamiltonians. As evident from the dominance of the
  peak at $n=0$, the amount of MT symmetry breaking
  largely stems from CM conserving terms. There is little difference between the
  degree of MT symmetry breaking in the different models. }
\label{fig:error66}
\end{figure}

We present the numerical results on MT symmetry breaking in our model Hamiltonians. Define the residual
\[H_{\text{res}}(n,n',l_1,l_2)=h(n,n',l_1,l_2)-H_p(l_1,l_2) \]
where, as before, $h(n,n',l_1,l_2)$ denotes the FCI interaction
Hamiltonian expressed in the Wannier basis. $H_{\text{res}}$ is the
part of $h(n,n',l_1,l_2)$ which does
not satisfy MT symmetry required by the PPs. Obviously, $H_{\text{res}}=0$ if $h$ is one of the PPs, since $h$ will then be equal to $H_p$. The quantity
\[\delta^2_n = \frac{\sum_{l_1l_2,m}|H_{\text{res}}(m,m+nL_y,l_1,l_2)|^2 }{\sum_{l_1l_2,m,m'}|h(m,m',l_1,l_2)|^2 }\]
allows us to track the origin of MT nonconservation. $\delta^2_0$
comprises the elements of $H_{\text{res}}$ satisfying $n=n'$. As defined in
Eq.~\ref{hp}, these are the elements which are independent of
$n$. $\delta^2_0$ hence represents the fraction of matrix elements that
are CM conserving but MT symmetry breaking. For $n\neq 0$,
$\delta^2_n$ represents MT nonconserving contributions that likewise
do not respect CM conservation. $\delta^2_n$ is plotted in
Fig.~\ref{fig:error66} for various model Hamiltonians, for a system
size $L_x=L_y=6$. The results remain almost unchanged when $L_x$ and $L_y$
are varied as long as $L_x=L_y>3$.

From the enhanced peak at $n=0$, we conclude that most of the MT
symmetry breaking occurs when CM is conserved. This happens because
our maximally localized Wannier functions (WFs) are still mostly peaked at one
site. The subdominant contributions from $\delta_n^2$ for $n=\pm 1$
can be attributed to the finite tails of the WFs one site away from their center of mass. Indeed, $\delta_n^2$ becomes exponentially small for $|n|>1$. While the overall extent of MT symmetry breaking originates from the nonuniformity of the Berry curvature, its relative contribution to $\delta_n^2$ for different $n$ is dictated by the localization properties of the WFs.

\subsection{Discussion on MT symmetry breaking}
The decomposition of the FCI Hamiltonian into pseudopotentials is
only exact in the thermodynamic LLL limit of zero
bandwidth and homogeneous
Berry curvature. For the generic model,
the FCI Hamiltonian can only be partly decomposed into pseudopotentials, which we then discuss along general FQHE pseudopotentials on the cylinder or torus. From our calculations, the
deviations are significant, suggesting that at least for the spectrum
above the elementary low energy quasiparticle regime,
there is no clear similarity between FCI and
FQH systems. However, entanglement signatures of incompressible liquid phases, such as the entanglement spectrum~\cite{li-08prl010504} with
the emergence of an entanglement gap~\cite{thomale-10prl180502}, show
strong similarities of FCI ground states to their FQH analogues, even
in terms of the counting rule of low-lying
states~\cite{regnault-11prx021014,bernevig-12prb075128}. This is
astonishing from the viewpoint of PPs, as the FCI and FQH models at
the bare level
could only possibly agree to the extent of the PP decomposable
components of the FCI Hamiltonians.

Such an apparent discrepancy between analyses at the Hamiltonian and entanglement measure levels can be interpreted as a consequence of a renormalization group flow. As high energy modes are integrated out, the low energy physics of FCIs supposedly flows towards FQHE type scenarios, with reemergent symmetries
such as magnetic translation group which is conserved in the FQHE but
broken in the FCI at a bare level. Recast into pseudopotentials, it suggests
that the PP non-decomposable part of the FCI Hamiltonian at the
bare level should
decrease upon renormalization, while the ratios of pseudopotentials of
the PP-decomposable part of the bare interactions might deviate from the PP
ratios in the low energy theory. This implies that even if two FCI
Hamiltonians have similar PP ratios at the bare level, they can
still differ considerably in their low energy description, and hence
their propensity to host FQHE-type incompressible states
(Fig.~\ref{fig:error66}).
This interpretation
is consistent with the common theme from FQHE numerical studies that the data quality of entanglement spectra and its characterization of the bulk and edge mode properties is not significantly correlated to the spectral sharpness of the Hamiltonian spectrum, and
partly anticipates the energy spectral flow~\cite{thomale-10prl180502}. Ultimately, only the joint confirmation of both entanglement and
Hamiltonian measures will justify true evidence for a fractional
topologically ordered phase in the FCI models.
From a low energy perspective, the seemingly clean finding from entanglement measures
might not yet rule out that the
inhomogeneous berry curvature induces a flow to a liquid
different from FQHE, as may also be seen by hints such that the hierarchy
liquid construction
cannot be established for the FCIs as in the FQHE
case~\cite{andreihierarchy}. From the perspective of energetics, the effective pseudopotential weights of the FCI
models in a low-energy theory are likely to be strongly
modified due to "integrating out'' the PP
non-decomposable part of the bare model, which can also provide an
explanation for the parameter trends of the stability of FCI phases as
a function of system parameters~\cite{adolfo}.

\section{Many-body pseudopotentials}
\label{sec:mbpp}

\subsection{Construction of many-body PP trial Hamiltonians}
While we have only explored two-body interactions so far,
PP expansions are also well-suited for many-body interactions. From the established knowledge in
FQHE systems, it follows that various
interesting FQAH liquids are located in the nullspace of certain many-body
PPs. In theory, we can construct many-body FCI
Hamiltonians that exhibit Pfaffian, Read-Rezayi etc. groundstates from
such
PPs~\cite{greiter-91prl3205,Read-99prb8084,simon-07prb075318,mcs}.

The first task is to generalize Haldane's PPs for FCI
models to more than two-body interactions~\cite{PhysRevB.75.195306}. For two particles in the
LLL and a translationally invariant potential $V(k)=\int
e^{-ik\cdot r} V(r)dr$, the component projected on the $m$th Legendre
component is given by Eq.~\ref{Vm}
\begin{equation*}
V^{m}=4\pi l_B^2\int\frac{d^2k}{(2\pi)^2}e^{-l_B^2k^2}L_m(l_B^2k^2)V(k)
\end{equation*}
so that the $m$th pseudopotential (with $V^m\propto \delta_{mn}$) is given by
$U^m(k)=V_0L_m(l_B^2k^2)$. As before, $V_0$ is a constant with units
of energy. In the plane limit, the $U^m$s form an orthogonal basis
which one can use to expand a generic potential profile.

When an interaction involves more than two particles, additional
complications arise. To begin with, there are different ways of choosing to
assign relative distance variables, or,  angular momentum. (For
two-body interactions, there is a unique assignment, as one degree of
freedom drops out due the CM conservation.)
 When
there are $3$ particles, only one degree of freedom is eliminated due to CM
conservation. As such, an ambiguity remains in choosing the many-body
analog of relative angular momentum. This ambiguity is mathematically
manifest when one tries to generalize Eq.~\ref{Vm}. In the case
of $3$-body interactions, there will be integrals over both momenta
$k_1$ and $k_2$ in the above expression, and
one has  to chose the new expression to involve $L_m(k_1^2)$, $L_m(k_2^2)$, $L_m((k_1-k_2)^2)$, or a combination of these.

This formal ambiguity similar to coupling multiple angular momenta does not induce physical complications as we
formulate generalized Haldane pseudopotentials (GHPs).
Its detailed first-principle derivation can be found in Appendix~\ref{nbody}. We
constrain ourselves to the application of GHPs to the total relative
angular momentum $N$-body PP Hamiltonians:
\begin{equation}
U^m(k)= L_m\left(\frac{k^2 l_B^2N}{2 (N-1)}\right).
\label{l2new}
\end{equation}
Here, $U^m(k)$ is the N-body interaction potential that has a total relative
angular momentum of $m$, with $k$ being the momentum conjugate to the total relative coordinate $w$ defined by
\begin{equation}
w= \frac{1}{N-1}\sum_{n=1}^{N-1}(z_n-z_N)=\frac{\sum_i^{N-1}z_i}{N-1}-z_N,
\label{nbodyw}
\end{equation}
where $z_i=x_i-iy_i$ are the complex coordinates of the $N$
particles.

In real space, the pseudopotential $U^m(w)\propto \int dk e^{ik\cdot
  w}U^m(k) $ depends explicitly on the positions of each of the $N$
particles. If we select one of the $N$ particles, the total relative
angular momentum is the sum of the relative angular momenta of the
other $N-1$ particles relative to the first one. Indeed, we see from
Eq.~\ref{nbodyw} that $w$ represents the relative seperation between
particle $N$ and the CM of the rest of the particles.
Note the appearance of the factor $\frac{N}{2(N-1)}$, which is essential
in obtaining the correct expressions for the PPs. (It will be derived in detail in Appendix~\ref{nbody}.)

The next step is to express $U^m(k)$ in the LLL Landau gauge
basis. Since the latter is the analogue of the Wannier basis, once we
have done so, we are able to read off the Wannier basis matrix
elements of a many-body FCI Hamiltonian exhibiting e.g. Pfaffian or
other more exotic groundstates. Here we shall perform this explicitly
for $N=3$ bosonic PPs. The computation with fermions or more bodies is conceptually similar.
According to Eq.~\ref{l2new}, we rescale the magnetic length $l^2$ by $\frac{3}{4}l^2$. Hence, $U^m$ becomes $V_0L_m(\frac{3}{4}k^2)$, where $k$ is the momentum conjugate to the total relative coordinate $((r_1-r_2)+(r_3-r_2))/2$. In the basis of LLL Landau gauge eigenfunctions,
\begin{widetext}
\begin{eqnarray}
&&U^m_{n_1n_2n_3n_4n_5n_6} \nonumber \\
&\propto&\sum_\sigma\int d^2r_1 d^2r_2d^2r_3
\psi^\dagger_{n_1}(r_1)\psi^\dagger_{n_2}(r_2)\psi^\dagger_{n_3}(r_3)L_m\left(\frac{3}{4}l_B^2
  \nabla^2_{[(r_1+r_3)/2-r_2]}\right)\psi ^{\phantom{\dagger}}_{n_4}(r_3) \psi ^{\phantom{\dagger}}_{n_5}(r_2) \psi ^{\phantom{\dagger}}_{n_6}(r_1)
\nonumber \\
&\propto&\sum_\sigma\int \frac{d^2qd^2p}{(2\pi)^4}\int \prod_i^3d^2r_i
L_m\left(\frac{3}{4}(p-q)^2 l_B^2\right)e^{i q\cdot (r_1-r_2)}e^{i
  p\cdot
  (r_2-r_3)}\psi^\dagger_{n_1}(r_1)\psi^\dagger_{n_2}(r_2)\psi^\dagger_{n_3}(r_3)\psi ^{\phantom{\dagger}}_{n_4}(r_3)\psi ^{\phantom{\dagger}}_{n_5}(r_2)\psi ^{\phantom{\dagger}}_{n_6}(r_1)
.\nonumber \\
\label{pseudopotmaim}
\end{eqnarray}
\end{widetext}
The $\sum_\sigma$ sum refers to a symmetric (antisymmetric) sum over all permutations
$\sigma$ assuming the particles are bosons (fermions). We have $k=q-p$ because
\begin{eqnarray}
e^{i q\cdot (r_1-r_2)}e^{i p\cdot
  (r_2-r_3)}&=&e^{i(q-p)\cdot((r_1+r_3)/2-r_2)}e^{i(p+q)\cdot(r_1-r_3)/2}
\nonumber \\
&=&e^{i(q-p)\cdot w_2}e^{i(p+q)\cdot w_3},
\end{eqnarray}
where $w_2$ and $w_3$ are linear combinations of the original coordinates whose roles will be further expounded in Appendix~\ref{nbody}. Here, it is sufficient to understand that $k$ should be the momentum conjugate to $w_2$, the total relative angular momentum. As before,
$\psi_n(r)=\frac{1}{\sqrt{\sqrt{\pi}Ll_B}}e^{i\frac{\kappa}{l_B}ny}e^{-\frac{(x-\kappa
    n l_B)^2}{2l_B^2}}$ and $\kappa=\frac{2\pi l_B}{L_y}$ is a dimensionless ratio that is small in the limit of large magnetic fields.

The integrals related to~\eqref{pseudopotmaim} can be simplified to a convenient form. As its
computation is instructive for the generalization to more
complicated cases, we explicate it in Appendix~\ref{3body}. The 3-body bosonic PP $U^0$ which hosts the Pfaffian ground state is given by
\[ U^0 \propto \sum_R \hat{b}_R^\dagger \hat{b}_R ^{\phantom{\dagger}}, \]
where
\begin{eqnarray}
&&\hat{b}_R\nonumber \\
&=&\sum_{n_i=3 R \text{mod} N}\left[\sum_{\sum s_i=0}
  e^{-\frac{\kappa^2}{2}\sum_i(R-(n_i+Ns_i))^2} \right]
c_{n_1}c_{n_2}c_{n_3} \nonumber \\
&=&\sum_{n_1+n_2+n_3=3 R \text{mod} N}\left[\sum_{s,t} e^{-\frac{\kappa^2}{3}W_{st}} \right] c_{n_1}c_{n_2}c_{n_3},
\end{eqnarray}
with $W_{st}=\sum_i {n'}_{i}^{2} -\sum_{i<j}n'_i n'_j$, $n'_1=n_1+sN$, $n'_2=n_2+tN$, and $n'_3=n_3-N(s+t)$. The $N=L_xL_y$ periodicity originates from the properties of the Wannier basis on the torus.
This result has been previously obtained in
Ref.~\onlinecite{dunghai2006}. However, the GHP formalism here can
also generate higher PPs of multi-body interactions, and this result is just its simplest case. Note that the summation constraint $n_1+n_2+n_3=3 R \text{mod} N$ can also be implemented as constraints over $s,t$, as in the 2-body case described in Ref.~\onlinecite{seidel-05prl266405}.

We see that $U^0$ is positive semidefinite. Since the Pfaffian state $|\text{Pf}\rangle$ resides in its kernel~\cite{greiter-91prl3205,dunghai2006}, we have
\[ \langle \text{Pf} |U^0|\text{Pf} \rangle = 0\]
as well as
\[ b^R|\text{Pf}\rangle =0. \]
As a consequence, the Pfaffian state is not just the ground state of
$U^0$ but is also annihilated by all $b_R$'s. Analogous to the
two-body case, $U^0$ consists of all three-body processes that
conserve the CM $R$. The physical positions of the
particles relative to their CM is described by $R-(n_i+Ns_i)$, where
$n_i$ is the real space basis index before enforcing the periodicity. When $\kappa$
is sufficiently small, contributions from periodic images can be
discarded, and we are left with a simpler expression with
$|R-(n_i+Ns_i)|<N=L_xL_y$.

Using our setup, we can likewise calculate higher PPs. $U^1$ vanishes due to the symmetric exchange statistics of the bosons. From Appendix~\ref{3body}, we have
\[ U^2 \propto \sum_R \hat{b}_R^\dagger \hat{b}_R ^{\phantom{\dagger}}, \]
where
\begin{widetext}
\begin{equation}
\hat{b}_R=\sum_{n_1+n_2+n_3 =3 R \text{mod} N}\left[\sum_{s,t} \left(1-\frac{2\kappa^2}{3}W_{st}\right)e^{-\frac{\kappa^2W_{st}}{3}} \right] c_{n_1}c_{n_2}c_{n_3},
\end{equation}
with $W_{st}=\sum_i {n'}_i^2 -\sum_{i<j}n'_in'_j$, $n'_1=n_1+sN$, $n'_2=n_2+tN$, and $n'_3=n_3-N(s+t)$ as before.
\end{widetext}

\subsection{Discussion}

While generic FCI models will only contain two-body interactions at the bare level,
our many-body pseudopotentials will be useful for the construction of trial Hamiltonians which exhibit new exotic groundstates. This situation is similar for
the FQHE case. There, the many-body PPs are primarily used to provide
trial Hamiltonians for which various FQHE states are exact null
modes. As stated before in Section III, the exact null mode property
is tied to the clustering property which is enfored by the $N$-body
trial Hamiltonian as $N$ particles approach each other. This will be
an interesting analogous application for the FCI scenario. By use of
our many-body PPs, various trial states can be realized in a FCI
model, including fractional Abelian states such as the Laughlin
series, fractional non-Abelian states such as the Read-Rezayi series, but
also even more exotic states such as a FCI Gaffnian
state~\cite{PhysRevB.38.3636,greiterbull,simon-07prb075317}. For the Pfaffian FCI state, for example, it is interesting to
study the interpolation of the exact FCI Hamiltonian to a generic FCI
model and see whether adiabatic connectivity can be reached both at
the level of Hamiltonians and entanglement spectra~\cite{thomale-10prl180502}.

Many-body FCI pseudopotentials will also be useful in studying
interactions beyond the effective single-band level. For instance, they are generically generated when interband scattering is considered.
Moreover, following the construction introduced in
Ref.~\onlinecite{lee-04prl096401} for the hardcore potential, we can
map all these states to effectively one-dimensional models (see also Appendix~\ref{nbody}
and~\ref{3body}), and even higher dimensional generalizations~\cite{PhysRevB.81.115123}, of
featureless Mott insulators with exotic ground state and quasiparticle
properties, which already are worth studying in their own right.

The use of many-body pseudopotential both as trial Hamiltonians and as
effective Hamiltonians opens up new branches of research. For example,
it will be interesting to apply the pseudopotential formalism to the model discussed in the recent work Ref.~\onlinecite{wangsheng}, where convincing numerical evidence for a stable $\nu=1$ non-Abelian state is presented. For our present work, we shall satisfy ourselves with developing the pseudopotential formalism.

\section{Conclusion}
\label{sec:concl}
We have developed a pseudopotential formalism for fractional Chern
insulators. Starting from the FCI Wannier state representation, we
have employed the FQHE formalism on the cylinder and torus to define
two-body as well as many-body PPs. We have decomposed bare FCI models
into PPs and find that generic situations give rise to
PP-decomposable and PP-nondecomposable parts of the FCI Hamiltonian,
and discussed their interplay with partial breaking of magnetic
translation group, which appears to be reemergent at low energies in
the FCI models. We have defined many-body PPs to establish a basis
for studying further unconventional FCI phases via appropriately
designed trial Hamiltonians. We believe that it will be interesting to
employ the PP perspective to provide a complementary tool to
entanglement measures and to further develop our understanding of
fractional Chern phases, as well as to extend its applicability to Chern bands
of higher Chern
number~\cite{PhysRevB.84.241103,maissam,maxemil,yanggusun,liulauch} and two-dimensional fractional topological insulator models~\cite{PhysRevLett.103.196803,PhysRevB.84.140404,PhysRevB.84.165107,PhysRevB.85.165134}.

\begin{acknowledgements}
We thank E.~J.~Bergholtz, C.~Chamon, C.~M.~Jian,
T.~Neupert, Z.~Papic, S.~A.~Parameswaran, A.~Seidel, and D.~N.~Sheng for discussions. CH is supported by a scholarship from the Agency of Science, Technology and Research of Singapore. RT is supported by an SITP fellowship by Stanford University. XLQ is supported by the Packard foundation and the Alfred P. Sloan foundation.
\end{acknowledgements}

\appendix

\section{Equivalence of the Landau-level-projected  and Cartesian description on
  the cylinder}
\label{equi}

We reconcile the two expressions for the delta function potentials on
the cylinder appearing in
Refs.~\onlinecite{rezayi-94prb17199,lee-04prl096401}. The following
arguments are valid for generic $l_B$, which has been set to unity in
the following. We show that
\begin{equation} \langle 0 | \delta^2(r-r')|0\rangle =
  e^{-(r_p-r'_p)^2/2},
\end{equation}
where $|0\rangle\sim f(z) e^{-|z|^2/4}$ denotes the LLL wavefunction and $r_p=(x_p,y_p)$ are the guiding-center coordinates defined by
\[ x_p = :\partial_z +\frac{z}{2}:, \]
\[ y_p = :-i\partial_z+\frac{iz}{2}: \]
with $z=x-iy$. These are the LLL-projected coordinates since $z_p =
z$, $\bar z_p = :2\partial_z:$(see Ref.~\onlinecite{jainbook}).
Here, the normal-ordering symbols ":'' indicate that any derivative
within them does not act on the Gaussian factor $e^{-|z|^2/4}$ present
in LL wavefunctions. Note that $[x_p,y_p]=i$, i.e. the guiding-center coordinates do not commute.
Consider a general interaction \[ V(r-r')= \int \frac{d^2k}{(2\pi)^2}V(k)e^{ik\cdot (r-r')}.\]
Let us project it onto the LLL $|0\rangle = f(z)e^{-|z|^2/4}$, i.e.

\[ \langle 0 |V(r-r')|0\rangle = \int \frac{d^2k}{(2\pi)^2}V(k)\langle 0 |e^{ik\cdot (r-r')}|0\rangle. \]

The quantity in the angle brackets is evaluated as
\begin{eqnarray}
\langle 0 |e^{ik\cdot r}|0\rangle &=&\langle 0 |e^{i(k\bar z + \bar{k}
  z)/2}|0\rangle \nonumber \\
&=& \langle 0 | e^{\frac{i}{\sqrt{2}}(\bar{k} a + k a^\dagger)}
e^{\frac{i}{\sqrt{2}}(\bar{k} b^\dagger + k b)}|0\rangle \nonumber \\
&=& e^{-|k|^2/2}\langle 0| e^{\frac{i}{\sqrt{2}}(k
  a^\dagger)}e^{\frac{i}{\sqrt{2}}(\bar{k} a )}
e^{\frac{i}{\sqrt{2}}(\bar{k} b^\dagger + k b)}|0\rangle \nonumber \\
&=& e^{-|k|^2/2}\langle 0| e^{\frac{i}{\sqrt{2}}(\bar{k} b^\dagger + k b)}|0\rangle,
\end{eqnarray}
where $k=k_x-ik_y$ and $a,b,a^\dagger,b^\dagger$ are lowering and
raising operators of angular momentum and LL (see also Appendix~\ref{nbody}). Due to the specific form of $|0\rangle$, we also have
\begin{widetext}
\begin{eqnarray}
b |0\rangle &=& \frac{1}{\sqrt{2}}(\frac{\bar
  z}{2}+2\partial_z)f(z)e^{-|z|^2/4} =
\frac{1}{\sqrt{2}}:2\partial_z:f(z)e^{-|z|^2/4} =\frac{1}{\sqrt{2}}z_p
|0\rangle, \nonumber \\
b^\dagger |0\rangle &=& \frac{1}{\sqrt{2}}(\frac{ z}{2}-2\partial_{\bar z})f(z)e^{-|z|^2/4} =   \frac{1}{\sqrt{2}}:z:f(z)e^{-|z|^2/4} = \frac{1}{\sqrt{2}}\bar z_p |0\rangle.
\end{eqnarray}
\end{widetext}
Hence,
\begin{eqnarray}
\langle 0 |e^{ik\cdot r}|0\rangle &=& e^{-|k|^2/2}\langle 0|
e^{\frac{i}{2}(\bar{k} z_p + k \bar z)}|0\rangle \nonumber \\
&=& e^{-|k|^2/2}\langle 0| e^{ik\cdot r_p}|0\rangle.
\end{eqnarray}
If we start from a delta function in real space, $V(k)=1$, we obtain
\[ \langle 0 |\delta^2(r-r')|0\rangle = \int \frac{d^2k}{(2\pi)^2} e^{-|k|^2/2}e^{ik\cdot r_p} = e^{-(r_p-r'_p)^2/2}.\]
Higher PPs in the LLL will thus be of the form
$L_m(\nabla^2_p) e^{-(r_p-r'_p)^2/2}$. To summarize, the LLL projection leads to two types of modifications. Firstly, the delta function is replaced by a Gaussian
in guiding-center coordinates. Secondly, the derivative will also be
taken with respect to guiding-center coordinates.

\section{First-principle derivation of Generalized Haldane Pseudopotentials}
\label{nbody}

We generalize the Haldane pseudopotentials to $N>2$ bodied interactions,
i.e. GHPs,
for fractionally filled Chern bands. The main technical steps in
this appendix include (i)  a derivation of the GHP starting from
an original interaction Hamiltonian involving arbitrarily many bodies
 (ii) a clarification on how the total relative angular
momentum can be defined through an appropriate change of coordinates. The results of this
appendix will be directly utilized in Appendix~\ref{3body} for explicit calculations of the first $3$-body PPs , which
always take the factorized form $\sim \sum_n \hat{b}^\dagger_n \hat{b}_n$.

We start with the $N$-body Hamiltonian
\begin{eqnarray}
H&=& \frac{1}{2m}\sum_i^N (\vec{p}_i + \frac{e}{c}\vec{A}_i)^2 +
V(\vec{x}_1,\vec{x}_2,...,\vec{x}_N) \nonumber \\
&=& \frac{1}{2m}\sum_i^N \left(-i\frac{\partial}{\partial
    x_i}-\frac{eB}{c}\frac{y_i}{2}\right)^2
+\left(-i\frac{\partial}{\partial
    y_i}+\frac{eB}{c}\frac{x_i}{2}\right)^2 \nonumber \\
&&+ V(\vec{x}_1,\vec{x}_2,...,\vec{x}_N) \nonumber \\
&=&\frac{1}{2m}\sum_i^N \left[ -4l_B^2\frac{\partial^2}{\partial
    z_i \partial \bar{z}_i}+\frac{1}{4l^2}|z_i|^2 +
  (\bar{z}_i\frac{\partial}{\partial
    \bar{z}_i}-z_i\frac{\partial}{\partial z_i})\right ]\nonumber \\
&&+V(\vec{x}_1,\vec{x}_2,...,\vec{x}_N),
\label{symmham}
\end{eqnarray}
where $z_i=x_i-iy_i$, $\vec{A}_i=B(-y_i,x_i,0)/2$ and
$l_B=\sqrt{\frac{\hbar c}{eB}}$. The symmetric gauge has been used so
that the Hamiltonian eigenstates will be conveniently labelled by
angular momentum. However, the results that follow will be
gauge-invariant. For now, we will not make any assumption about the
form of the interaction potential $V$.

Haldane's original procedure was to first seperate this Hamiltonian into a CM part and relative part, and then project the relative part into different angular momentum sectors~\cite{haldane83prl605}. The same will be done here, but with $N$ particles instead of two. Define a change of coordinate
\[w_i = R_{ij} z_j, \]
with $w_1=(z_1+z_2+\dots + z_N)/N$, i.e. the CM coordinate. Any part of the resultant Hamiltonian depending only on $w_1$ will not affect our PP expansion.

\subsection{Allowed coordinate transforms $R_{ij}$}
It turns out that we are not free to choose the rest of $R_{ij}$ arbitrarily. If we want to have a well-defined angular momentum decomposition in terms of the new variables $\{ w_i \}$, we will need to ensure that the resultant Hamiltonian is of the same form as the last line of  Eq.~\ref{symmham}. This is because the kinetic term in~\eqref{symmham} can be written as
\begin{eqnarray}
H_{kin}&=&\frac{1}{2m}\sum_i^N \left[ -4l_B^2\frac{\partial^2}{\partial
    z_i \partial \bar{z}_i}+\frac{1}{4l^2}|z_i|^2 +
  (\bar{z}_i\frac{\partial}{\partial
    \bar{z}_i}-z_i\frac{\partial}{\partial z_i})\right ] \nonumber \\
&=& \hbar \omega \sum_i( b^\dagger_i b ^{\phantom{\dagger}}_i+\frac{1}{2})+\hbar
\omega(a^\dagger_i a ^{\phantom{\dagger}}_i -b^\dagger_i b ^{\phantom{\dagger}}_i) \nonumber \\
&=&\hbar\omega\sum_i (a^\dagger_i a ^{\phantom{\dagger}}_i+\frac{1}{2}),
\end{eqnarray}
with eigenstates labeled by angular momentum $m$ and LL index $n$:
\begin{equation}
|n,m\rangle = \frac{(b^\dagger)^{m+n}}{\sqrt{(m+n)!}}\frac{(a^\dagger)^n}{\sqrt{n!}}|0,0\rangle.
\end{equation}
This can be seen from how the angular momentum operator exist as part
of the kinetic single-particle Hamiltonian. With the second-quantized operators defined by (particle index $i$ suppressed) \[a=\frac{1}{\sqrt{2}}\left(\frac{z}{2l_B}+2l_B\frac{\partial}{\partial\bar{z}}\right),\]
\[b=\frac{1}{\sqrt{2}}\left(\frac{\bar{z}}{2l_B}+2l_B\frac{\partial}{\partial z}\right),\]
the angular momentum operator is  \[L=\hbar \left(\bar{z}\frac{\partial}{\partial\bar{z}}-z\frac{\partial}{\partial z}\right ) = \hbar(a^\dagger a - b^\dagger b).\]
To expand in terms of angular momentum eigenstates, the coordinate transform $R_{ij}$ must leave the form of each term of the last line of Eq.~\ref{symmham} invariant, i.e. $|z_i|^2$ must transform into a sum of similar quadratic terms, etc.

Denoting $z=(z_1,z_2,...,z_N)^T$, $\partial z=(\frac{\partial}{\partial z_1},\frac{\partial}{\partial z_2},...,\frac{\partial}{\partial z_N})^T$, and likewise for the $w^i$s, the various terms transform as
\begin{equation} \sum_i \frac{\partial^2}{\partial z_i \partial \bar{z}_i}=\partial \bar{z}^T \partial z = \partial \bar{w}^T[ R R^T] \partial w,
\end{equation}
\begin{equation} \sum_i |z_i|^2= \bar{z}^T z =  \bar{w}^T[ (R^{-1})^T(R^{-1})]  w=\bar{w}^T[R R^T]^{-1} w,
\end{equation}
\begin{equation} \sum_i z_i\frac{\partial}{\partial z_i }=z^T \partial z = w^T[ (R^{-1})^T R^T] \partial w = w^T \partial w.
\end{equation}
The Hamiltonian retains the same form if $R R^T$ is diagonal. If we regard $R$ as a rotation matrix, we see that this condition is satisfied whenever $R$ maps an orthogonal basis to another orthogonal basis. Hence an allowed $R$ consists of mutually orthogonal rows. As a simple example, the $R$ matrix for 2 particles satisfies the condition
\[
\left( \begin{array}{cc}
1/2& 1/2  \\
1 & -1\end{array} \right),
\]
according to the CM coordinate $w_1=(z_1+z_2)/2$ and the relative coordinate $w_2=z_2-z_1$.

\subsection{The explicit form of the Hamiltonian transformed into total relative coordinates}

The next step is to explicitly find the coefficients of the transformed Hamiltonian. Since we are interested in a PP expansion in the total angular momentum, we define the total relative coordinate
\begin{equation}
w_2=
\frac{1}{N-1}\sum_{n=1}^{N-1}(z_n-z_N)=\frac{\sum_i^{N-1}z_i}{N-1}-z_N
.
\end{equation}
The other coordinates can be arbitrarily defined as long as they are orthogonal to $w_2$ and $w_1=\frac{1}{N}\sum_i^{N}z_i$.
With this choice, the diagonal elements of $RR^T$ are
\[ \lambda_1 = N, \lambda_2 = \frac{N}{N-1} , \dots \]
Hence the kinetic part of the Hamiltonian becomes
\begin{eqnarray}
H_{\text{kin}}&=&\frac{1}{2m}\sum_i^N \left[
  -4l_B^2\lambda_i\frac{\partial^2}{\partial w_i \partial
    \bar{w}_i}+\frac{1}{4l_B^2\lambda_i}|w_i|^2 \right. \nonumber \\
&& \left. +
  (\bar{w}_i\frac{\partial}{\partial
    \bar{w}_i}-w_i\frac{\partial}{\partial w_i})\right ] \nonumber \\
&=&\frac{1}{2m} \left[ -4l^2_{\text{rel}}\frac{\partial^2}{\partial
    w_2 \partial \bar{w}_2}+\frac{1}{4l^2_{\text{rel}}}|w_2|^2 \right. \nonumber \\
&& \left. + (\bar{w}_2\frac{\partial}{\partial \bar{w}_2}-w_2\frac{\partial}{\partial w_2})\right ]+\dots,
\end{eqnarray}
where $l_{\text{rel}}=l_B\sqrt{\lambda_2}=l_B\sqrt{\frac{N}{N-1}}$ is the effective "magnetic length" for the total relative coordinate. Only the terms corresponding to the total relative coordinate are shown in the second line.
In general, the diagonal elements of $RR^T$ $\lambda_1,\lambda_2,\lambda_3,...,\lambda_N$ define a set of effective magnetic lengths $l_i=l_B\sqrt{\lambda_i}$.

\subsection{Derivation of the N-body pseudopotential}

We are now set up to find $\langle m_1,...,m_N|V| m_1,...,m_N
\rangle$, the projection of an interaction potential
$V(w_1,w_2,...,w_N)$ onto the angular momentum sector in the LLL. This
projection is of course dependent on $w=Rz$. For the $w_1$ and $w_2$
previously defined as the CM and total relative coordinates, $m_1$ and
$m_2$ correspond to the CM angular momentum and total relative angular
momentum, respectively. To evaluate this matrix element, we Fourier transform to shift the coordinate dependencies onto an universal exponential factor:
\begin{widetext}
\begin{eqnarray}
V^{m_1m_2\dots m_N}
&=&(4\pi)^N\langle m_1,...,m_N|V(w_1,...,w_N)|m_1,...,m_N\rangle \nonumber \\
&=&(4\pi)^N \langle
m_1,...,m_N|\left(\prod_j^N\int\frac{d^2(k_jl_j)}{(2\pi)^2}\right)V(k_1,...,k_N)\prod_j^N
e^{i k_j\cdot w_j}|m_1,...,m_N\rangle \nonumber \\
&=& (4\pi )^N\left(\prod_j^N\int\frac{d^2(k_jl_j)}{(2\pi)^2}\right)\prod_j^N \langle
m_j|V(k_1,...,k_N) e^{i k_j\cdot w_j}|m_j\rangle \nonumber \\
&=& (4\pi )^N\left(\prod_j^N\int\frac{d^2(k_jl_j)}{(2\pi)^2}\right)V(k_1,...,k_N)\prod_j^N \langle m_j| e^{i k_j\cdot w_j}|m_j\rangle.
\end{eqnarray}
The momentum-space potential $V$ in the last line can be taken out of
the expectation value since the momenta labeled by $k_j$s are regarded as
complex numbers.
For each $j$,
\begin{eqnarray}
\langle m_j| e^{i k\cdot w_j}|m_j\rangle &=&\langle m_j| e^{i
  \frac{l_j}{2}(k\bar{z}+\bar{k}z)}|m_j\rangle =\langle m_j| e^{\frac{il_j}{\sqrt{2}}(\bar{k}a_j +k a^\dagger_j)
}e^{\frac{il_j}{\sqrt{2}}(\bar{k}b^\dagger_j +k b_j) }|m_j\rangle
\nonumber \\
&=&e^{-|k|^2l_j^2/4}\langle m_j| e^{i\frac{l_j}{\sqrt{2}}ka^\dagger_j
}e^{i\frac{l_j}{\sqrt{2}}\bar{k} a_j
}e^{i\frac{l_j}{\sqrt{2}}(\bar{k}b^\dagger_j +k b_j) }|m_j\rangle
=e^{-|k|^2l_j^2/2}\langle
m_j|e^{\frac{il_j}{\sqrt{2}}\bar{k}b^\dagger_j
}e^{\frac{il_j}{\sqrt{2}}k b_j } |m_j\rangle \nonumber \\
&=&e^{-|k|^2l_j^2/2}\sum_{s=0}\frac{1}{(s!)^2}\langle m_j|
\left(\frac{il_j\bar{k}b^\dagger}{\sqrt{2}}\right)^s\left(\frac{il_j k
    b}{\sqrt{2}}\right)^s|m_j\rangle = e^{-|k|^2l_j^2/2}\sum_{s=0}\frac{m!}{(s!)^2(m-s)!}\left(\frac{-l_j^2
    |k|^2}{2}\right)^s \nonumber \\
&=&e^{-|k|^2l_j^2/2}L_m\left(\frac{l_j^2|k|^2}{2}\right).
\label{mvm}
\end{eqnarray}
The terms containing the $a_j$ and $a_j^\dagger$ operators in the
third line reduce to unity because the states are already defined to
be in the LLL. Use has been made of the Baker-Campbell-Hausdorff
formula in producing the factors of $e^{-|k|^2l_j^2/4}$.
\end{widetext}

The LLL projected pseudopotential component reads
\begin{eqnarray}
&&V^{m_1m_2\dots m_N} \nonumber \\
&=&\langle m_1,...,m_N|V(w_1,w_2,...,w_N)|m_1,...,m_N\rangle \nonumber
\\
&=& \prod_j^N\int\frac{d^2(k_jl_j)}{\pi}e^{-|k_j|^2l_j^2/2}L_{m_j}\left(\frac{l_j^2|k_j|^2}{2}\right)V(k_1,...,k_N),\nonumber \\
\label{vmm}
\end{eqnarray}
where $V(k_1,...,k_N)$ is the Fourier transform of $V(w_1,...,w_N)$.
We focus on the $w_2$ degree of freedom. Let $|m\rangle$ denote the
state where $m_2=m$ and all other $m_i=0$, i.e. the state with total
relative angular momentum $m$. Then the $m$th PP component for a translationally invariant interaction is
\begin{eqnarray}
V^m&=&
\left(\prod_{j=2}^N\int\frac{d^2(k_jl_j)}{\pi}e^{-|k_j|^2l_j^2/2}\right)\nonumber
\\
&&\times L_m\left(\frac{N l_B^2|k_j|^2}{2(N-1)}\right)V(k_2,k_3,...,k_N).
\label{vm}
\end{eqnarray}
The integral over $k_1$ has been omitted since $V$ does not depend on the CM coordinate $w_1$. Also, the rest of the Laguerre polynomial factors have disappeared since $L_0=1$.
We still need to define $V^{m_2,...m_N}(k_2,k_3,...,k_N)$ such that it
corresponds to a PP component \[\langle
m_1,...,m_N|V(w_1,w_2,...,w_N)|m_1,...,m_N\rangle \] that is nonzero
only at the simultaneous set of angular momenta $m_2,...,m_N$. This is
done by exploiting the orthonormality relation
$\int_0^{\infty}2qe^{-q^2} L_s(q^2)L_t(q^2)dq=\delta_{st}$. Switching
to polar coordinates, we see that the functional form of the
pseudopotential is given by
\begin{equation}
U^{m_2,...m_N}=V_0\prod_{j=2}^N L_{m_j}\left(\frac{k_j^2 l_j^2}{2}\right),
\end{equation}
where $V_0$ is again a constant with units of energy. If we want a PP that has no angular momentum on the spurious degrees of freedom $w_3,w_4,...$, we find
\begin{equation}
U^m(k)= V_0L_m\left(\frac{k^2 l_B^2N}{2 (N-1)}\right).
\label{l2}
\end{equation}
This reproduces the familiar result $U^m(k)= L_m\left(k^2
  l_B^2\right)$ for $N=2$. For $N=3$, $U^m(k)= L_m\left(\frac{3}{4}k^2
  l_B^2\right)$ where $m$ is the total relative angular momentum
characterized by $w_2=\frac{1}{2}((z_1-z_3) + (z_2-z_3))$. The latter
will be used extensively in the calculations of the next section.

\subsection{Discussion and Generalizations}
We can determine how the effective magnetic length $l_j$ should be
rescaled without assuming the explicit form $w_j$ defined in terms of the $z_i$s. By examining the diagonal elements of $R R^T$, we find that
\[l_i= l_B\sqrt{\sum_j^{N-1}R_{ij}^2+\sum_{j<k}^{N-1}R_{ij}R_{ik}} \]
for $i>1$. For $i=1$, $l_1=\sqrt{N}l_B$. Here, $w_1$ is the CM position
and $l_1$ is the effective magnetic length for the CM angular
momentum.

\eqref{vmm} can also be extended to cases beyond the LLL. There, the
$a$ and $a^\dagger$ terms in the third line of Eq.~\ref{mvm} will
not yield unity. If we consider the case where each particle occupies
a specific Landau Level, we will have to first calculate expressions
such as $e^{i\frac {l_j}{\sqrt{2}}ka^\dagger_j }e^{i\frac{l_j}{\sqrt{2}}\bar{k} a_j}$ before making the change of coordinates from $z$ to $w$. This is because the positions of the particles are indexed by $z$, not $w$.

To begin with, we rearrange the exponential factor in the Fourier transform $e^{i k_j\cdot w_j}=e^{i(k_jR_{ji})z_i}$ so that it depends explicitly on the $z_i$s, albeit with modified $k=(k_jR_{ji})$, summation implied. After some algebra, we find
\[\langle n'| e^{\frac{il_j}{\sqrt{2}}ka^\dagger_j }e^{\frac{il_j}{\sqrt{2}}\bar{k} a_j}|n\rangle=
\left( - \frac{i l_Bk}{\sqrt{2}}\right)^{n'-n}\sqrt{\frac{n!}{n'!}}L_n^{n'-n}\left(\frac{l_B^2|k^2|}{2}\right),\]
where $n$ and $n'$ denote the initial and final Landau levels of the
particle. When $n=n'$, we just have $L_n(k^2 l_B^2/2)$. With the
states being reduced to the LLL, we proceed as in~\eqref{vmm}, arriving
at the general formula for the PP between $N$ particles
initially at LLs $n_1,n_2,\dots,n_N$ mapping to LLs
$n'_1,n'_2,\dots,n'_N$, with angular momenta $m_1,m_2,\dots,m_N$
associated with the coordinates $w_i=R_{ij}z_j$:
\begin{widetext}
\begin{equation}
V_{n'_1\dots n'_N;n_1\dots n_N}^{m_1m_2\dots m_N}=
\langle  n'_1,...,n'_N;m_1,...,m_N|V(w_1,w_2,...,w_N)| n_1,...,n_N;m_1,...,m_N\rangle=  \nonumber\end{equation}
\begin{equation}
\left(\prod_j^N\int\frac{d^2(k_jl_j)}{\pi}e^{-|k_j|^2l_j^2/2}L_{m_j}\left(\frac{l_j^2|k_j|^2}{2}\right)\right)\prod_i^N
\left( \frac{-i l_B(R_{li}k_l)}{\sqrt{2}}\right)^{n'_i-n_i}\sqrt{\frac{n_i!}{n'_i!}}L_{n_i}^{n'_i-n_i}\left(\frac{l_B^2|(R_{li}k_l)^2|}{2}\right) V(k_1,...,k_N),
\label{grand}
\end{equation}
\end{widetext}
with the effective magnetic lengths $l_j$ as before.

If $V$ is translationally invariant, it does not depend on $w_1=(z_1+\dots+z_N)/N$, and the $k_1$ integral produces a delta function $\delta(k_1)$. Hence, as is usually the case, $k_1$ should be excluded from all sums in Eq.~\ref{grand}.
As an illustration for $N=2$ bodies with interaction independent of
the CM, only the $k_2$ integration survives. We have $R_{li}k_l=\pm k_2$
and $V=V(k_2)$. $m_1$, the CM angular momentum, is irrelevant for
the interaction, so $V_{m_1}=\delta_{m_1,0}$. We can also
deduce this result from the orthogonality of the Laguerre polynomials:
if we impose the further restriction $n_i=n'_i$ for all $i$, i.e. particles stay in their respective LLs, Eq.~\ref{grand} reduces to
\begin{widetext}
\begin{equation}
V_{n_1n_2;n_1n_2}^{m_2}\propto\int d^2k e^{-l_B^2k_2^2}L_m(l_B^2k_2^2)L_{n_1}\left(\frac{l_B^2k_2^2}{2}\right)L_{n_2}\left(\frac{l_B^2k_2^2}{2}\right)V(k_2).
\label{simple}
\end{equation}
\end{widetext}

\section{Derivation of 2-body Pseudopotential Hamiltonians on a Cylinder}
\label{2body}

Here, we present the details of the derivation of the two-body PPs $U^m$. We shall explicitly work through only the fermionic case, since the bosonic case can be analogously derived.
From Eq.~\ref{lag}, a special case of Eq.~\ref{simple}, the potential $U^m$ that is nonzero only for the relative angular momentum sector $m$ is given by $U^m(r-r')= V_0 L_m(-l_B^2\nabla^2)\delta^2(r-r')$, where $V_0$ is a constant with units of energy.  We find its LLL Landau gauge basis matrix elements $U^m_{n_1n_2n_3n_4}$ by projecting onto the basis wavefunctions $\psi_n(r)=\frac{1}{\sqrt{\sqrt{\pi}L_yl_B}}e^{i\frac{\kappa}{l_B}ny}e^{-\frac{(x-\kappa nl_B)^2}{2l_B^2}}$, where $L_y$ is the circumference of the cylinder:
\begin{widetext}
\begin{eqnarray}
U^m_{n_1n_2n_3n_4}&=&\frac{V_0}{4}\int d^2r
d^2r'\psi^\dagger_{n_1}(r)\psi^\dagger_{n_2}(r') L_m(-l_B^2
\nabla^2)\delta^2(r-r')\psi ^{\phantom{\dagger}}_{n_3}(r')\psi ^{\phantom{\dagger}}_{n_4}(r) + \text{antisymm} \nonumber \\
&=&\frac{V_0}{4}\int \frac{l_B^2 d^2q}{(2\pi)^2}\int d^2r d^2r' L_m(q^2l_B^2)e^{i q\cdot (r-r')}\psi^\dagger_{n_1}(r)\psi^\dagger_{n_2}(r')\psi ^{\phantom{\dagger}}_{n_3}(r')\psi ^{\phantom{\dagger}}_{n_4}(r) + \text{antisymm}.
\label{pseudopotmain}
\end{eqnarray}
\end{widetext}
Recall that $\kappa=\frac{2\pi l_B}{L_y}$ is a dimensionless ratio that
is small in the limit of large magnetic fields. The two types of MT
symmetry constraints mentioned in Section V are manifest in the above
expression as (i) the CM conservation which
corresponds to $n_1+n_2=n_3+n_4$, and (ii) one-dimensional translation
symmetry, which is the invariance of $U^m$ under $n_i\rightarrow n_i
+a$, where $a$ is an integer and $i=1,2,3,4$. (A similar observation
has been independently made in Ref.~\onlinecite{gunnar}.)
CM conservation must be present because the $\int dy$ and $\int dy'$ integrals produce delta functions of the form
\[ \delta\left(\frac{2\pi}{L_y}(n_4-n_1)+q_y\right)\] and
\[ \delta\left(\frac{2\pi}{L_y}(n_3-n_2)-q_y\right).\]
The CM conservation condition $n_1+n_2=n_3+n_4$, i.e. $n=n'$, appears after we combine these two delta functions. Since the CM of each LLL wavefunction occurs along $x=\kappa l_Bn\propto n$, we see that the CM of the particles must indeed be equal before and after a two-body hopping.

\begin{figure}[t]
\begin{minipage}{0.99\linewidth}
\includegraphics[width=0.47\linewidth]{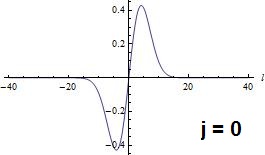}
\includegraphics[width=0.47\linewidth]{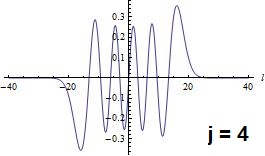}
\includegraphics[width=0.47\linewidth]{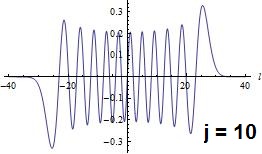}
\includegraphics[width=0.47\linewidth]{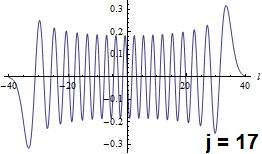}
\end{minipage}
\caption{Graphs of $b^{2j+1}$ from Eq.~\ref{bj} for $j=0,4,10,17$ with $\kappa=\frac{1}{6}$. As $j$ increases, the
  main region of contribution of $b^{2j+1}$ shifts in the direction of
  larger $|l|$.}
\label{bpic}
\end{figure}

We explicitly resolve the MT symmetry via the translation
$n_i\rightarrow n_i+1$, as Fourier terms cancel upon
$\psi_i(r)\rightarrow \psi_i\left(r-\frac{2\pi l_B^2}{L_y}\right)$. This
is exactly a magnetic translation under which the system in a magnetic
field $B=\frac{\hbar c}{el_B^2}$ is expected to be invariant.
To continue the calculation, reduce $U^m$ to the $\int d^2q$ integral
\begin{widetext}
\begin{eqnarray}
U^m_{l_1l_2}&=& \frac{V^0}{4\pi L_y^2l_B^4}\frac{L_y}{2\pi}\pi l_B^2\int d^2q
\delta(q_y+\frac{2\pi}{L_y}(l_1-l_2))L_m(l_B^2q^2)e^{-\frac{l_B^2q^2_x}{2}}e^{i\kappa
  l_B(l_1+l_2)q_x}e^{-\frac{\kappa^2 (l_1-l_2)^2}{2}}+\text{antisymm}
\nonumber \\
&=&\frac{V^0}{4\pi L_y^2l_B^2}\frac{L_y}{2\pi}\pi l_B^2\int d^2q
\delta(q_y+\frac{2\pi}{L_y}(l_1-l_2))L_m(l_B^2q^2)e^{-\frac{l_B^2q^2}{2}}e^{i\kappa
  l_B(l_1+l_2)q_x}+\text{antisymm}.
\label{qint}
\end{eqnarray}
The second line follows from the fact that $q_y$ is constrained to be
$q_y=\frac{2\pi}{L_y}(l_2-l_1)=\frac{\kappa}{l_B}(l_2-l_1)$. Note that
$n$ completely disappears from the expression, as expected from MT
symmetry. A closed form expression for $U^m$ is given by
\begin{eqnarray}
&& U^m_{l_1l_2}=\nonumber \\
&&g\kappa^3\sum_{p=0}^m\frac{(-1)^{p+1}m!}{p!(m-p)!}\sum_{j=0}^p\sum_{r=0}^{j/2}\frac{\Gamma(p-j+r+1/2)2^{m+j-r-3}(i\kappa)^{2(j-r-1)}}{(p-j)!(2r)!(j-2r)!\sqrt{\pi}(l_1l_2)^{2r-j}}\left[(l_1+l_2)^{2r}-(-1)^j(l_1-l_2)^{2r}
\right]e^{-\kappa^2(l_1^2+l_2^2)}.\nonumber \\
\end{eqnarray}
\end{widetext}
The lengthy expression above can be factorized into the form
$U^m_{l_1l_2}=g\kappa^3 b^m_{l_1}b^m_{l_2}$, $g={4V_0}{(2\pi)^{3/2}}$, where
\begin{equation}
b^{2j+1}_l = le^{-\kappa^2l^2}\sum_{p=0}^j \frac{(-2)^{3p-j}(\kappa l)^{2p}\sqrt{(2j+1)!}}{(j-p)!(2p+1)!}.
\label{bj}
\end{equation}
This result can be proven by induction. The first few $b^{2j+1}$s are
\[ b^1_l=le^{-\kappa^2l^2},\]
\[ b^3_l=\frac{1}{\sqrt{3!}}(-3+4\kappa^2l^2)le^{-\kappa^2l^2},\]
\[ b^5_l=\frac{1}{\sqrt{5!}}(15-40\kappa^2l^2+16\kappa^4l^4)le^{-\kappa^2l^2},\]
\[ b^7_l=\frac{1}{\sqrt{7!}}(-105+420\kappa^2l^2-336\kappa^4l^4 +64\kappa^6l^6)le^{-\kappa^2l^2},\]
\[ b^{2j}_l=0.\]
Certain $b$'s are depicted in Fig.~\ref{bpic} for illustration. Note that the $U^m$ operators can always be decomposed into the
product of two $b^m$ operators that are $m$ degree polynomials of
$l_1$ and $l_2$. These polynomials have the physically relevant
property that (i) the $b^m_l$ of higher $m$ are "localized" at larger
values of $l$ and (ii)
that they are orthogonal in the limit of $\kappa\rightarrow 0$ before
we enforce the $L_xL_y$ periodicity in the $x$-direction of the cylinder. This is
further explained in Appendix~\ref{ortho}.

We can similarly calculate the bosonic PPs through Eq. \ref{qint}, but with terms symmetrized instead of antisymmetrized over $l_1$ and $l_2$. As before, the PPs can be written as $U^m_{l_1l_2}=g\kappa^3 c^m_{l_1}c^m_{l_2}$, $g={4V_0}{(2\pi)^{3/2}}$, but now with the $c^m$s taking the form

\[ c^0_l=e^{-\kappa^2l^2},\]
\[ c^2_l=\frac{1}{\sqrt{2!}}(-1+4\kappa^2l^2)le^{-\kappa^2l^2},\]
\[ c^4_l=\frac{1}{\sqrt{4!}}(3-24\kappa^2l^2+16\kappa^4l^4)e^{-\kappa^2l^2},\]
\[ c^6_l=\frac{1}{\sqrt{6!}}(-15+180\kappa^2l^2-240\kappa^4l^4 +64\kappa^6l^6)e^{-\kappa^2l^2},\]
\[ c^{2j+1}_l=0.\]

Note that both the bosonic and fermionic results can also be obtained via Gram-Schmidt Orthogonalization of the basis set comprising of even(odd) powers of $\kappa l$ with the inner product measure $e^{-2\kappa^2l^2}$. Indeed, the functional forms of their polynomial part can be uniquely determined by orthonormality requirements once we have obtained the form of their inner product measure in Eq. \ref{bj}.

\section{Orthogonality of $U^m$}
\label{ortho}

The FQH PPs $U^m$ are not systematically orthogonal once we place them
on a cylinder or torus. We will study the physical origin of this fact and
make quantitative estimates in this appendix. While
we are only concerned with quadratic cases of $L_x=L_y$ in this paper,
it will be instructive to investigate how deviations from orthogonality depend on general $L_x$ and $L_y$.

\subsection{From plane to cylinder}
When we compactify the plane into a cylinder, the notion of relative
angular momentum is no longer well-defined. In a fixed LL, the
relative angular momentum is proportional to the interparticle
distance which can only be meaningfully defined when the latter is much smaller than $L_y$. Recall that the effective $\kappa=\frac{2\pi\l_B}{L_y}=\frac{1}{L_y}$ after mapping to the FCI system. We thus expect orthogonality to occur only in the limit of large $L_y$, or, equivalently, small $\kappa$.
To proceed, we evaluate the overlap elements of two PPs $V_{\text{cyl}}^m$ and $V_{\text{cyl}}^n$, and show that the latter are orthogonal for sufficiently small $\kappa$.
\begin{widetext}
\begin{eqnarray}
\langle {U}_{\text{cyl}}^m,U_{\text{cyl}}^n \rangle&\propto & \sum_{ l_1
  l_2}U^m_{\text{cyl }  l_1, l_2}U^n_{\text{cyl }  l_1, l_2}
\nonumber \\
&\propto&\sum_{ l_1 l_2}\int d^2q\int d^2q'
\delta(q_y+\frac{2\pi}{L_y}(l_1-l_2))\delta(q'_y-q_y)L_m(l_B^2q^2)L_n(l_B^2q'^2)e^{-\frac{l_B^2(q^2+q'^2)}{2}}e^{i\kappa
  l_B(l_1+l_2)(q_x+q'_x)}
  \nonumber \\
&\propto &\sum_{ l_1+l_2, l_1-l_2}\int dq_y\int dq_x \int dq'_x
\delta(q_y+\frac{2\pi}{L_y}(l_1-l_2))L_m(l_B^2q^2)L_n(l_B^2q^2)e^{-\frac{l_B^2(q^2+q'^2)}{2}}e^{i\kappa
  l_B(l_1+l_2)(q_x+q'_x)}
 \nonumber \\
&\propto&\sum_{l_1-l_2}\int dq_y\int dq_x \int dq'_x
\delta(q_y+\frac{2\pi}{L_y}(l_1-l_2))L_m(l_B^2q^2)L_n(l_B^2q^2)e^{-l_B^2q^2}\delta(q_x+q'_x)
\nonumber \\
&\propto&\int d^2q
\left(\sum_{l_1-l_2}\delta(q_y+\frac{2\pi}{L_y}(l_1-l_2))\right)L_m(l_B^2q^2)L_n(l_B^2q^2)e^{-l_B^2q^2}
\nonumber \\
&\approx&\pi\int
2(ql_B)d(ql_B)\left(\sum_{l_1-l_2}\delta((q_yl_B)+\kappa(l_1-l_2))\right)L_m(l_B^2q^2)L_n(l_B^2q^2)e^{-l_B^2q^2}
\nonumber \\
&\rightarrow & \pi \delta_{mn}.
\end{eqnarray}
\end{widetext}
Two approximations have been made above. From the third last to second
last line, we replace the non-rotationally invariant integral over
$\int dq_x dq_y$ by the rotationally invariant integral $\pi\int
2qdq$. From the second last to the last line, the delta function sum
in the large parentheses was replaced by unity, i.e. taking the limit where the range of $l_1-l_2$ (and hence $q$) tends to infinity. These approximations become exact when the discrete $q_yl_B$ becomes a continuum, which occurs precisely when $\kappa\rightarrow 0$. Indeed, this agrees with the physical intuition that the relative angular momentum becomes a well-defined quantity when $L_y$ is large.
From the viewpoint of polynomial orthogonality, we see that the
orthogonality of the $U_{\text{cyl}}^m$ is respected as much as as the
integral over Laguerre polynomials is allowed to be made
continuous. Specifically, the orthogonality of the Laguerre
polynomials is exact only for the continuum $q^2=q^2_x+q^2_y$ and not
for discrete points specified by $\Delta l =l_1-l_2$ whose separations do not vanish unless $\kappa= 0$.

\subsection{From cylinder to torus}

The compactification of the cylinder into a torus introduces a periodicity in the $L_x$-direction.
This introduces periodic copies of $U^m_{\text{cyl}}$ in
$U^m_{\text{tor}}$, each displaced from another by $(L_xL_y,L_xL_y)$
or $(L_xL_y,-L_xL_y)$ sites in $l_1-l_2$ space. Nonorthogonality is
expected when there is significant overlap between these images. Let us obtain a bound by finding the conditions where $|U_{\text{cyl}}^m|$, whose explicit form is given by~\eqref{bj}, is not negligible, with $l_1,l_2$ being of the order of $L_xL_y$. At such values,
\begin{eqnarray}
 U^{m}_{\text{cyl} l_1 l_2} &\sim& e^{-2(\kappa L_xL_y)^2}(2^{3/2}\kappa L_xL_y)^{2m}\nonumber \\
 & =& e^{-2w^2}(2^{3/2}w)^{2m}\nonumber \\
 &=&f(w),
\end{eqnarray}
where $w=\kappa L_xL_y$. The function $f(w)$ exhibits a rapid Gaussian decay beyond $w\approx 2\sqrt{m}$. Hence we expect the PP $U^m$ to be nonorthogonal when
\[\kappa  < \frac{2\sqrt{m}}{L_xL_y}, \]
which implies that
\begin{equation} L_x <2\sqrt{m},
\label{orthobound}
\end{equation}
a bound well-verified by calculations in the following subsection.

\subsection{Numerical results for the orthonormality of the fermionic pseudopotentials}
\begin{figure}
\begin{minipage}{0.99\linewidth}
\includegraphics[width=0.8\linewidth]{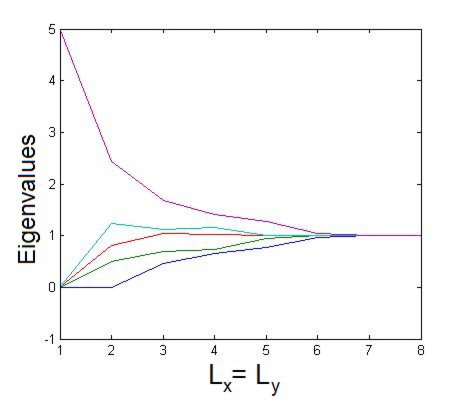}
\end{minipage}
\caption{(Color online) The eigenvalues of the overlap matrix of
  $U^1,U^3,U^5,U^7$ and $U^{9}$ as a function of the system size
  $L=L_x=L_y$. Indeed, we observe orthogonality when $L\geq 6$, in
  excellent agreement with the bound $2\sqrt{9}= 6$ from
  Eq.~\ref{orthobound}.}
\label{fig:phaseKM}
\end{figure}

The orthonormality of the $U^m$s can be studied quantitatively through
their overlap matrix defined by
\begin{equation}
M_{mn}=\langle U^m, U^n \rangle
\end{equation}
according to Eq.~\ref{over}. When the $U^m$s are orthonormal, it holds
$M_{mn}=\mathbb{I}$. If the $U^m$s are not orthonormal while they span an orthonormal basis, the
eigenvalues of $M_{mn}$ will still be unity since we can find a
unitary transformation where $M_{mn}$ is diagonal. When $U^m$s are
overcomplete, however, the spectrum of $M$ broadens and yields
eigenvalues deviating from this limit.
In Fig.~\ref{fig:phaseKM}, we plot the eigenvalues of the overlap
matrix for the first few PPs as a function of
$L$. Orthogonality is hence broken when the eigenvalues differ
from unity. Indeed, we observe that orthogonality improves with system
size, in agreement with the conclusions of the preceding
subsections.
\\

\section{Three-body bosonic Pseudopotential calculations in the LLL Landau gauge basis}
\label{3body}

Let $U^m$ be the bosonic 3-body CM-invariant potential which is
nonzero only in the sector of total relative angular momentum
$m$. Such $3$-body $U^m$s will be useful as a basis through which arbitrary CM conserving potentials can be expanded. In this appendix, we will show the detailed derivation of $U^m$ in the basis of LLL Landau gauge eigenfunctions, starting from Eq.~\ref{pseudopotmain}.

According to Eq.~\ref{l2}, we have to replace the magnetic length
$l_B^2$ by $\frac{3}{4}l_B^2$. This, however, causes no difference when
$m=0$, since $L_0=1$, so the results in Refs. \onlinecite{lee-04prl096401} and
\onlinecite{dunghai2006} for $U^0$ could have been obtained without
this modification along our GHPs.

After performing the $\int d^2r_i$ integrations according to Eq.~\ref{pseudopotmain}, we arrive at
\begin{widetext}
\begin{equation}
U^m_{n_1n_2n_3n_4n_5n_6}\propto\sum_\sigma\int \frac{d^2qd^2p}{(2\pi)^4} I(q,n_1,n_6)I(p-q,n_2,n_5)I(-p,n_3,n_4)L_m\left(\frac{3}{4}(p-q)^2 l_B^2\right),
\end{equation}
where
\[ I(q,n,n')=2\pi\delta\left(q_y+\frac{\kappa}{l_B}(n-n')\right)e^{-l_B^2q^2_x/4}e^{i\kappa l_B(n+n')q_x/2}e^{-\kappa^2(n-n')^2/4}.\]
Here, $\sum_\sigma=\frac{1}{(3!)^2}\sum_{perm(n_1n_2n_3)}\text{sgn}(perm)^f\sum_{perm(n_6n_5n_4)}\text{sgn}(perm)^f$ where $f=1$ for fermions and $f=0$ for bosons. $perm$ refers to the permutations of the $n_i$s while $sgn(perm)=\pm 1$ depending on whether the permutation is even or odd.
Up to now, the expression is easily generalizable to any number of fermions or bosons by increasing the number of integrals and $I(q,n,n')$s. If we further restrict ourselves to the 3-body bosonic case, we can simplify it to
\begin{eqnarray}
&&U^m_{n_1n_2n_3n_4n_5n_6} \nonumber \\
&\propto& e^{-\kappa ^2([(n_1-n_6)^2+(n_2-n_5)^2+(n_3-n_4)^2]/4}\sum_\sigma
\int \frac{d^2q_xd^2p_x}{(2\pi)^4}L_m\left(\frac{3}{4}(p-q)^2
  l_B^2\right)e^{-l_B^2(p^2+q^2+(p-q)^2)/4}e^{i\kappa l_B N_1 q/2}e^{i\kappa l_B N_2 p/2}
\nonumber \\
&\propto& \sum_\sigma P_m(n_1,n_2,n_3,n_4,n_5,n_6;\kappa ^2)
\frac{4\pi}{\sqrt{3}}e^{-\frac{\kappa ^2}{6}(N_1^2+N_1N_2+N_2^2)}, \nonumber \\
\end{eqnarray}
\end{widetext}
where $N_1=n_1+n_6-n_2-n_5$,
$N_2=n_2+n_5-n_3-n_4$. $P_m(n_1,n_2,n_3,n_4,n_5,n_6;\kappa ^2)$ is a
potentially complicated polynomial whose form depends on $m$. We
define the conserved CM "Landau level wavefunction index" $R$ via $3 \text{R \text{mod} N}=n_1+n_2+n_3=n_6+n_5+n_4$. The existence of $R$ is a conseqence of the total CM conservation of $U^m$.

\subsection{m=0 for bosons on a torus}
In this case, $L_0=1$, and $P_0(n_1,n_2,n_3,n_4,n_5,n_6;\kappa^2)=1$. Since each of the $n_i$'s is also defined modulo $N$ (but constrained to sum to $3R$ as shown above), we make the replacement
\begin{eqnarray}
&& e^{-\frac{\kappa ^2}{2}((R-n_1)^2+(R-n_2)^2+(R-n_3)^2)}\rightarrow
\nonumber \\
&& \sum_{s,t} e^{-\frac{\kappa ^2}{2}((R-(n_1+Ns))^2+(R-(n_2+Nt))^2+(R-(n_3-N(t+s)))^2)}, \nonumber\\
\end{eqnarray}
and likewise for the identical factor involving $n_4,n_5,n_6$. Hence the $U^m$ factorizes into a product of nonlocal operators $\hat{b}_R$:
\[ U^0 \propto \sum_R \hat{b}_R^\dagger \hat{b}_R^{\phantom{\dagger}}, \]
where
\begin{eqnarray}
&&b_R= \nonumber \\
&&\sum_{\sum_i n_i=3 R \text{mod} N}\left[\sum_{\sum s_i=0}
  e^{-\frac{\kappa ^2}{2}\sum_i(R-(n_i+Ns_i))^2} \right]
c_{n_1}c_{n_2}c_{n_3} \nonumber \\
&=&\sum_{n_1+n_2+n_3=3 R \text{mod} N}\left[\sum_{s,t} e^{-\frac{\kappa ^2}{3}W_{st}} \right] c_{n_1}c_{n_2}c_{n_3},
\end{eqnarray}
with $W_{st}=\sum_i {n'}_i^2 -\sum_{i<j}n'_i n'_j$, $n'_1=n_1+sN$, $n'_2=n_2+tN$ and $n'_3=n_3-N(s+t)$.
This result is identical to that in Ref. \onlinecite{dunghai2006}.

\subsection{m=1 for bosons on a torus}

In this case, the 1st Laguerre polynomial gives a factor $1-\frac{3}{4}(\kappa ^2(n_2-n_5)^2+p_x^2)$, and $\sum_\sigma P_1$ evaluates to
\begin{eqnarray}
\sum_\sigma P_1&=& \sum_\sigma 3\kappa ^2(n_2-R)(n_5-R) \nonumber \\
&=&
3\kappa ^2\left(\frac{n_2+n_1+n_3}{3}-R\right)\left(\frac{n_4+n_5+n_6}{3}-R\right)
\nonumber \\
&=&0
\end{eqnarray}
Hence we have
\begin{equation}
U^1=0.
\end{equation}
This agrees with the result from Ref. \onlinecite{simon2007}, in that
there is no PP of total relative angular momentum $m=1$
for bosons. This is because the CM invariance of the interaction
precludes any symmetric wave function of total degree 1.

\subsection{m=2 for bosons on a torus}
After some algebra,
\[P_2=\frac{1}{2}(1-3\kappa ^2(n_2-R)^2)(1-3\kappa ^2(n_5-R)^2).\]
Since
\begin{eqnarray}
\sum_\sigma (n_2-R)^2=&=& \sum_\sigma( n_2^2-2n_2 R + R^2) \nonumber \\
&=& \frac{1}{3}\sum_{i=1}^3 n_i^2 - 2R^2 + R^2\nonumber \\
&=& \frac{2}{9}(\sum_i n_i^2 -\sum_{i<j}n_in_j),
\end{eqnarray}
we have
\[ U^2 \propto \sum_R \hat{b}_R^\dagger \hat{b}_R ^{\phantom{\dagger}}, \]
where
\begin{equation}
\hat{b}_R=\sum_{\sum_i n_i=3 R \text{mod} N}\left[\sum_{s,t} \left(1-\frac{2\kappa ^2}{3}W_{st}\right)e^{-\frac{\kappa ^2}{3}W_{st}} \right] c_{n_1}c_{n_2}c_{n_3}
\end{equation}
with $W_{st}=\sum_i {n'}_i^2 -\sum_{i<j}n'_i n'_j$, $n'_1=n_1+sN$, $n'_2=n_2+tN$, and $n'_3=n_3-N(s+t)$ as before.

\subsection{m=3 for bosons on a torus}

This is zero for 2 bosons, but not for 3 bosons~\cite{simon2007}. Indeed,

\[\sum_\sigma P_3= \sum_\sigma \frac{9}{2}\kappa ^6(n_2-R)^3(n_5-R)^3 = \frac{9}{2}\kappa ^6\prod_{i=1}^6(n_i-R).\]
Hence
\[ U^3 \propto \sum_R \hat{b}_R^\dagger \hat{b}_R ^{\phantom{\dagger}}, \]
where
\begin{widetext}
\begin{equation}
\hat{b}_R=\frac{-3\kappa ^3}{\sqrt{2}}\sum_{n_1+n_2+n_3=3 R \text{mod} N}\left[\sum_{s,t} (n_1+sN-R)(n_2+tN-R)(n3-(s+t)N-R)e^{-\frac{\kappa ^2}{3}W_{st}} \right] c_{n_1}c_{n_2}c_{n_3}.
\end{equation}
\end{widetext}

\subsection{m=4 for bosons on a torus}
After the smoke clears, we find
\[ U^4 \propto \sum_R \hat{b}_R^\dagger \hat{b}_R^{\phantom{\dagger}}, \]
where
\begin{eqnarray}
&&\hat{b}_R=\nonumber \\
&&\sum_{\sum_i n_i=3 R \text{mod} N}\left[\sum_{s,t}
  \left(1-\frac{2\kappa ^2W_{st}}{3}+\frac{\kappa
      ^4W^2_{st}}{9}\right)e^{-\frac{\kappa ^2}{3}W_{st}} \right]
c_{n_1}c_{n_2}c_{n_3}, \nonumber \\
\end{eqnarray}
with $W_{st}=\sum_i {n'}_i^2 -\sum_{i<j} n'_i n'_j$, $n'_1=n_1+sN$,
$n'_2=n_2+tN$, and $n'_3=n_3-N(s+t)$ as before.


\begin{thebibliography}{10}
\bibitem{remarksun}
the original CB model\cite{sun-11prl236803} distinguishes between two types of
  NNN hoppings $t'_1$ and $t'_2$, but the constraint $t'=t'_1=-t'_2$ is
  sufficient to produce a band of maximum flatness.
\end{thebibliography}

\end{document}